\documentclass[prb,twocolumn,superscriptaddress,longbibliography]{revtex4-1}

\usepackage{amsmath}
\usepackage{amssymb}
\usepackage{amsthm}
\usepackage{amsfonts}
\usepackage{listings}
\usepackage{threeparttable}
\usepackage{enumerate}
\usepackage{latexsym}
\usepackage{color}
\usepackage{bm}
\UseRawInputEncoding
\usepackage{hyperref}
\hypersetup{
 pdfnewwindow=true, colorlinks=true,
 linkcolor=blue, anchorcolor=blue,
 citecolor=blue, filecolor=blue,
 menucolor=blue, urlcolor=blue}

\usepackage{psfrag}

\usepackage{bm}
\usepackage{graphicx}

\newcommand{\webirvsp}{\href{https://github.com/zjwang11/irvsp}{\ttfamily irvsp}}
\newcommand{\webchecktopmat}{\href{https://www.cryst.ehu.es/cryst/checktopologicalmagmat}{\ttfamily Check Topological Mat}}

\newcommand{\beq}{\begin{equation}}
\newcommand{\eneq}{\end{equation}}
\def\bk{{\bf k}}
\def\bp{{\bf p}}

\def\ie{{\it i.e.},\ }
\def\eg{{\it e.g.},\ }

\begin{document}

\tolerance 10000

\newcommand{\vk}{{\bf k}}

\draft

\title{%
Topological insulators in the NaCaBi family with large SOC gaps
}
\author{Dexi Shao}
\affiliation{Beijing National Laboratory for Condensed Matter Physics,
and Institute of Physics, Chinese Academy of Sciences, Beijing 100190, China}
\author{Zhaopeng Guo}
\affiliation{Beijing National Laboratory for Condensed Matter Physics,
and Institute of Physics, Chinese Academy of Sciences, Beijing 100190, China}
\affiliation{National Laboratory of Solid State Microstructures and
School of Physics, Nanjing University, Nanjing 210093, China}

\author{Xianxin Wu}
\affiliation{Max-Planck-Institut f\"{u}r Festk\"{o}rperforschung, Heisenbergstrasse 1, D-70569 Stuttgart, Germany}
\affiliation{Beijing National Laboratory for Condensed Matter Physics,
and Institute of Physics, Chinese Academy of Sciences, Beijing 100190, China}

\author{Simin Nie}
\affiliation{Department of Materials Science and Engineering, Stanford University, Stanford, California 94305, USA}

\author{Jian Sun}
\email[]{Corresponding author. Email: jiansun@nju.edu.cn}
\affiliation{National Laboratory of Solid State Microstructures and
School of Physics, Nanjing University, Nanjing 210093, China}
\affiliation{Collaborative Innovation Center of Advanced Microstructures, Nanjing 210093, China}

\author{Hongming Weng}
\affiliation{Beijing National Laboratory for Condensed Matter Physics,
and Institute of Physics, Chinese Academy of Sciences, Beijing 100190, China}
\affiliation{University of Chinese Academy of Sciences, Beijing 100049, China}

\author{Zhijun Wang}
\email[]{Corresponding author. Email: wzj@iphy.ac.cn}
\affiliation{Beijing National Laboratory for Condensed Matter Physics,
and Institute of Physics, Chinese Academy of Sciences, Beijing 100190, China}
\affiliation{University of Chinese Academy of Sciences, Beijing 100049, China}

\date{\today}

\begin{abstract}
By means of first-principles calculations and crystal structure searching techniques, we predict that a new NaCaBi family crystallized into the ZrBeSi-type structure (\ie $P6_{3}/mmc$) 
are strong topological insulators (STIs).
Taking $P6_{3}/mmc$ NaCaBi as an example,
the calculated band structure indicates that there is a band inversion between two opposite-parity bands at the $\Gamma$ point.
In contrast to the well-known Bi$_2$Se$_3$ family, the band inversion in the NaCaBi family has already occured even without spin-orbit coupling (SOC), giving rise to a nodal ring surrounding $\Gamma$ in the $k_z=0$ plane (protected by $M_z$ symmetry).
With time reversal symmetry $(\cal T)$ and inversion symmetry $(\cal I)$, the spinless nodal-line metallic phase protected by $[{\cal TI}]^2=1$ is the weak-SOC limit of the spinful topological insulating phase. Upon including SOC, the nodal ring is gapped, driving the system into a STI.
Besides inversion symmetry, the nontrivial topology of NaCaBi can also be indicated by $\bar{6}$ symmetry.
More surprisingly, the SOC-induced band gap in NaCaBi is about 0.34 eV, which is larger than the energy scale of room temperature. Four other compounds (KBaBi, KSrBi, RbBaBi and RbSrBi) in the family are stable at ambient pressure, both in thermodynamics and lattice dynamics, even though the gaps of them are smaller than that of NaCaBi. Thus, they provide good platforms to study topological states both in theory and experiments.

\end{abstract}

\maketitle

\begin{figure*}[!t]
\begin{center}
\includegraphics[width=0.98\textwidth]{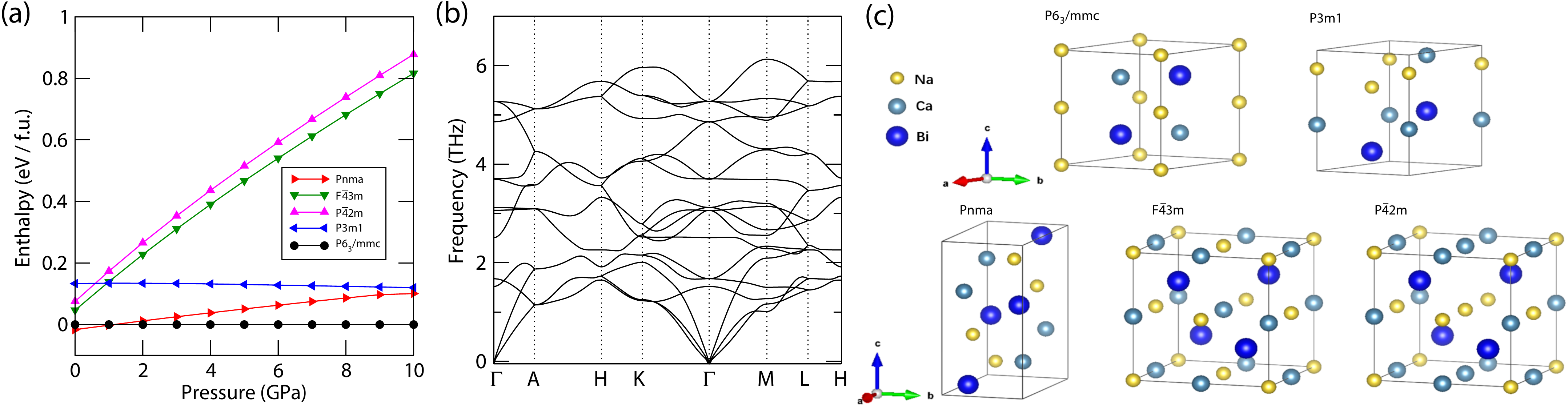}
\caption{%
  (a) Calculated enthalpy relative to that of $P6_{3}/mmc$ phase vs. pressure. (b) Phonon dispersions for the $P6_{3}/mmc$ phase at the ambient pressure, which justify the lattice dynamical stability of this phase.
  (c) Five crystal structures crystallized into the $P6_{3}/mmc$, $P3m1$, $Pnma$, $F\bar{4}3m$ and $P\bar{4}2m$ phases of stoichiometric NaCaBi near the convex hull from the formation enthalpies.
}
\label{fig:str}
\end{center}
\end{figure*}

\section{INTRODUCTION}
Topological insulators (TIs) have been studied broadly and deeply in both theory and experiments, since the two-dimensional quantum spin Hall effect (2D TI) was proposed in 2005~\cite{Kane2005}. By generalizing the concept of TIs, the three-dimensional (3D) TIs were predicted in the Bi$_2$Se$_3$ family~\cite{Bi2X3-zhj2009}, which were verified in the experiments later~\cite{Hsieh2009,Bi2X3-Hasan2009,Bi2Te3-Shen2009}. TIs embody a new quantum state characterized by a topological invariant rather than a spontaneously broken symmetry. The nontrivial topology of TIs is well defined by the wave functions of occupied bands, as long as there is a continuous energy gap between the conduction bands and valence bands in the whole Brillouin zone (BZ).
Then, the $\cal T$-invariant TIs are generalized to topological crystalline insulators (TCIs)~\cite{TCI-FuLiang-2011,TCI-SG-2013,TCI-TSC-2014,Wang2016Hourglass,Ma2017}, in which the lattice symmetries play important roles. Generally, symmetry eigenvalues and irreducible representations (irreps) can indicate topological property. For example, the well-known Fu-Kane parity indices are widely used for centrosymmetric TIs~\cite{Fu-Kane-2007}, and the recently established symmetry indicators and topological quantum chemistry are very efficient to diagnose the nontrivial topology of TIs and TCIs~\cite{wzj2019,Bradlyn2018,Bradlyn2019,Ashvin-SI-2017,Songzd-TCI-2018,Bradlyn2017,ZhangTT2019,TangF2019,Vergniory2019}.
Although many candidates for TIs/TCIs have been proposed, the TIs protected by sixfold rotoinversion symmetry [\ie the sixfold rotoinversion $\bar 6~(\equiv {\cal I} C_6)$ is a sixfold rotation ($C_6$) followed by inversion ($\cal I$)] are rarely reported.

The 3D TIs have attracted tremendous attention because their surface states possess odd spin-momentum locked Dirac cones, thus, the backscattering channels of the surface states are suppressed, leading to dissipationless electronic transport~\cite{Hsieh2009,Bi2X3-zhj2009,TIs-Hasan2010, TIs-FuLiang2011}. In proximity to an $s$-wave superconductor (SC), the topological surface states can acquire an effective $p$-wave superconducting paring, resulting in a 2D topological SC~\cite{PhysRevLett.100.096407}.
In addition, the TIs can be tuned into many interesting topological states. For instance, the quantum anomalous Hall (QAH) insulator is found in the Cr-doped Bi$_2$Te$_3$~\cite{Yu2010,Chang2013}.
Furthermore, 
the compound MnBi$_2$Te$_4$ is predicted to be an antiferromagnetic TI at zero magnetic field and a Weyl semimetal under the out-of-plane magnetic field~\cite{xuyong2019}. Recently, the QAH effect has been observed in few-layer thin films~\cite{nwaa089}.
Moreover, the unconventional SC has been proposed and some experimental evidences were revealed in Cu-doped Bi$_2$Se$_3$~\cite{PhysRevLett.104.057001,NP2010}. The strong evidences of Majorana zero-energy mode within a vortex in the TI/SC heterostructure -- Bi$_{2}$Te$_{3}$/NbSe$_{2}$ -- have been proposed recently~\cite{PhysRevLett.114.017001}.
Therefore, the TI candidates are of great interest, especially those with relatively large band gaps, which not only benefit the the future applications of spintronics and quantum computation from the dissipationless surface states, but also provide good platforms to study the interplay between the topological phase and symmetry-breaking phases. 
Though many TI candidates and corresponding interesting properties have been proposed~\cite{Bi2X3-zhj2009,TIs-Hasan2010, TIs-FuLiang2011,Wu2016,Nie2018}, TIs with relatively large band gaps and clear 2D Dirac-cone surface states as the Bi$_2$Se$_3$ family are rare.

In this work, using crystal structure searching techniques and first-principles calculations, we predict that NaCaBi can be stable in the space group (SG) of $P6_{3}/mmc$ at pressures up to 10 GPa from both the thermal dynamics and the lattice dynamics.
Calculations of the band structures indicate that without SOC, $P6_{3}/mmc$ NaCaBi is a topological semimetal with a nodal ring surrounding the $\Gamma$ point in the $k_{z}=0$ plane. 
Upon including SOC, the crossing points of the nodal line are all gapped and the system becomes a STI.
We find that the nontrivial topology can be indicated not only by inversion symmetry, but also by the $\bar{6}$ symmetry.
Moreover, it is exciting that the topologically nontrivial band gap is about 0.34 eV due to strong SOC, which is even larger than the energy scale of room temperature.
At last, four other dynamically stable XYBi compounds with similar electronic structures and STI natures are proposed in Section~4 of Appendix. 

\section{Calculations and Results}
\subsection{The crystal structures of stoichiometric NaCaBi}
We used crystal structure prediction techniques integrated in USPEX~\cite{Oganov2006,Artem2011,Lyakhov2013} and the machine learning accelerated crystal structure search method~\cite{Gao2018} 
to find the best candidates of NaCaBi under pressure. From the phase diagrams based on convex hull analysis of formation enthalpies for the Na-Ca-Bi ternary system at 0 GPa and 10 GPa shown in Fig.~\ref{fig:convex}, we find stoichiometric NaCaBi crystallizes in $Pnma$ phase at 0 GPa and $P6_{3}/mmc$ phase at 10 GPa
, respectively. We extract several structures of stoichiometric NaCaBi near the two stable phases from the formation enthalpy, which are in SGs of $P6_{3}/mmc$, $P3m1$, $Pnma$, $F\bar{4}3m$ and $P\bar{4}2m$, respectively.
The enthalpy-pressure ($\triangle H-P$) curves between these phases plotted in Fig.~\ref{fig:str}(a) exhibit the best ones from our structural predictions at pressures up to 10 GPa.
At ambient pressure, the most stable structure of NaCaBi is found to be $Pnma$ phase, and the enthalpy of $P6_{3}/mmc$ phase is just ~0.17 eV/$f.u.$ higher than that of $Pnma$ phase. Our calculations reveal that $P6_{3}/mmc$ phase becomes preferred at pressures higher than 1 GPa.

As shown in Fig.~\ref{fig:str}(b) and Fig.~\ref{fig:xyz}, there is no phonon mode with negative frequency in the phonon spectra of $P6_{3}/mmc$ NaCaBi under pressures from 0 GPa to 10 GPa, which indicates that the metastable $P6_{3}/mmc$ phase could be quenched to ambient pressure if they can be synthesized at higher pressure.
These results are in accordance with the earlier work~\cite{Zhang2012,Peng2018}.
As shown in Fig.~\ref{fig:md}(c), calculations of the ab initio molecular dynamic (AIMD) simulations at ambient pressure and T = 600 K indicates that there is no structural collapse after 10 ps (10000 steps).
In addition, motivated by earlier works~\cite{Levy2010,Shao2017}, how the parameters evolves under pressure, strain potential and the possible substrates are also discussed in Section~8 of Appendix.
As shown in Fig.~\ref{fig:xyz} and Table~\ref{table:xyz}, we also list many other XYBi compounds in the $P6_{3}/mmc$ phase which are dynamically stable at low pressures. Besides NaCaBi, other four XYBi compounds (\ie KBaBi, KSrBi, RbBaBi, and RbSrBi) have similar band structures, thus the same STI nature with NaCaBi (see Section~4 of the Appendix).
Different from NaCaBi, the $P6_{3}/mmc$ phase of KBaBi, KSrBi, RbBaBi and RbSrBi are always energetically preferred from 0 GPa to 10 GPa, as shown in Fig.~\ref{fig:pnma}. Moreover, the NaCaBi system possesses the largest inverted band gap, as shown in Table~\ref{table:bandgap}. In the following, we will focus on NaCaBi with $P6_{3}/mmc$ phase in the main text.

\begin{table}[!b]
\begin{ruledtabular}
\caption{Band gaps of the NaCaBi family at 0 GPa.}
   \begin{tabular}{p{1.4cm} c c c c c}%
          \textbf{XYBi}         & NaCaBi     &  KBaBi  &  KSrBi   & RbBaBi    &  RbSrBi\\\hline
          \textbf{Gaps~(eV)}     & 0.34      &  0.22   &  0.26    & 0.20      &   0.25\\
  \end{tabular}
  \label{table:bandgap}
\end{ruledtabular}
\end{table}

\subsection{The band structures of $P6_{3}/mmc$ NaCaBi}
\begin{figure}
\begin{center}
\includegraphics[width=0.45\textwidth]{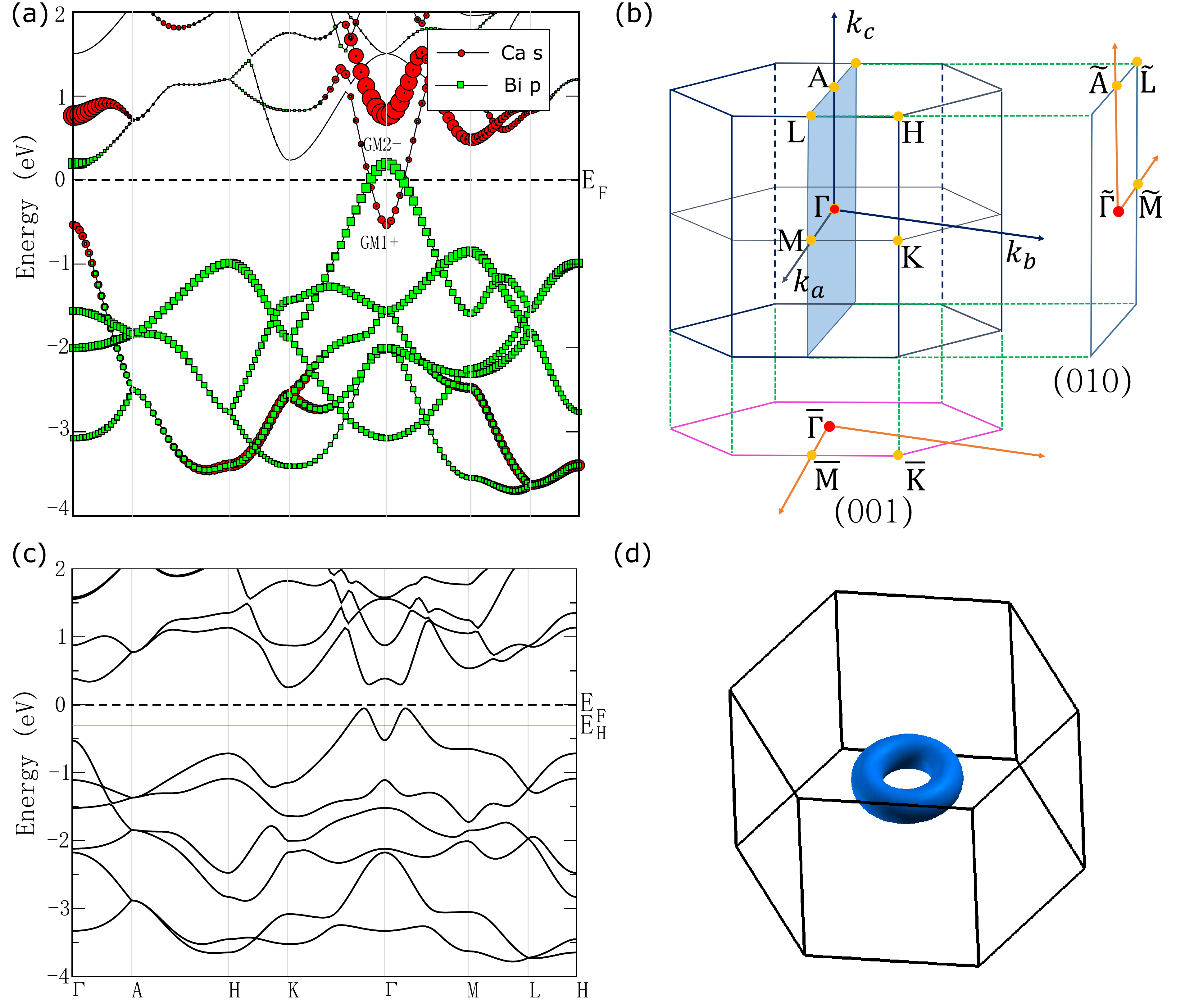}
\caption{%
  (a) Orbital-resolved band structures of $P6_{3}/mmc$ NaCaBi at 0 GPa without SOC.
  (b) The BZ of $P6_{3}/mmc$ NaCaBi, where the high-symmetry $\vec{k}$ points are marked out.
  (c) Band structures of $P6_{3}/mmc$ NaCaBi at 0 GPa with SOC. The red line corresponds to the 0.06 hole doping case (\ie E$_H$-E$_F$=-0.25 eV).
  (d) The constant-energy contour surface at E$_H$ is a torus.
}
\label{fig:2}
\end{center}
\end{figure}

From the band structures of $P6_{3}/mmc$ NaCaBi without SOC shown in Fig.~\ref{fig:2}(a), we can find two band crossings along K--$\Gamma$ and $\Gamma$--M at $E_F$.
The orbital-resolved band structures in Fig.~\ref{fig:2}(a) indicate that the valence band maximum is contributed by the Bi-$p$ states, while the conduction band minimum is from the Ca-$s$ states (which also hybridize with the Bi-$s$, Na-$s$ and \text{Ca-$d$} states).
Calculations of the irreps~\cite{gao2021} indicate that the two bands belong to GM$1+$  and  GM$2-$ irreps respectively [Fig.~\ref{fig:2}(a)], in the convention of the Bilbao Crystallographic Server notations at $\Gamma$ (GM).
The parities of the occupied bands at eight time-reversal invariant momenta (TRIMs) are given in Table~\ref{table:parity}.
Thus, we conclude that the band inversion appears between opposite-parity bands around $\Gamma$ without SOC, which guarantees the existence of nodal line(s) in the spinless systems with $\cal T$ and $\cal I$. Furthermore, this nodal ring surrounding $\Gamma$ lies in the $k_z=0$ plane 
due to the presence of $M_z$ symmetry (\ie the two inverted bands have different $M_z$ eigenvalues).

The spinful topological insulating phase can be considered as a (strong-)SOC limit of the spinless nodal-line metallic phase protected by $[{\cal TI}]^2=1$~\cite{wzj2019}.
As shown in Fig.~\ref{fig:2}(c), we can find that SOC induces a visible band gap between the two inverted bands around $\Gamma$, but it does not change the ordering of energy bands at $\Gamma$. The nodal ring is fully gapped, thus, the NaCaBi becomes a STI. More surprisingly, the SOC-induced band gap is as large as 0.34 eV, which is comparable with thermal energy of room temperature. The $P6_{3}/mmc$ phase of NaCaBi family compounds provides good platforms to study topological states both in theory and experiments.
Considering the possible underestimation of the band gap by Perdew-Burke-Ernzerhof (PBE) method~\cite{GGA-PBE1996}, the band inversion can be further confirmed by the calculations using hybrid Heyd-Scuseria-Ernzerhof (HSE) functional~\cite{HSE2006}. The results shown in Section~5 of the Appendix indicate that the SOC-induced band gap is almost unchanged, although the inverted gap at $\Gamma$ (\ie $E_{g}\equiv|E_{\text{GM1+}}-E_{\text{GM}2-}|$) becomes smaller.

Symmetry eigenvalues (or irreps) play key roles in identifying TIs/TCIs in solids~\cite{Bradlyn2019,Ashvin-SI-2017,Songzd-TCI-2018}. As we know, SG 194 belongs to a $\mathbb Z_{12}$ classification. Using
\webirvsp~\cite{gao2021} and \webchecktopmat~\cite{Vergniory2019}, all symmetry indicators are calculated to be
$z_{2w,i=1,2,3}=0;~z_{4}=3;~z_{6m,0}=5;~z'_{12}=11$.
By performing the symmetry analysis, we find that there are two essential symmetries indicating the nontrivial topology in this system. One is the inversion symmetry, while the other is the $\bar{6}$ symmetry. In other words, even though the inversion is broken when the SG is reduced to SG 174 (P$\bar{6}$), the nontrivial topology can be still indicated by $z_{3m,0}=1$ and $z_{3m,\pi}=0$. With $M_z\equiv [\bar{6}]^3$, these indicate that the mirror Chern numbers of $k_z=0$ and $k_z=\pi$ planes are 1 and 0, respectively. Thus, this system has to be a 3D STI.

\begin{table}[!t]
\begin{ruledtabular}
\caption{The product of the parities for all the occupied bands
at the eight TRIMs for the $P6_{3}/mmc$ phase of NaCaBi.}
   \begin{tabular}{c c c c c c}%
          \textbf{TRIM}         & $\Gamma$  &  3M   &  A    &  3L    & Product\\\hline
           Parity               &    +      &  --   &  --   &  --    & --\\
  \end{tabular}
  \label{table:parity}
\end{ruledtabular}
\end{table}

\subsection{The surface states of $P6_{3}/mmc$ NaCaBi}
Exotic topological surface states serve as significant fingerprints to identify various topological phases. Based on the tight-binding model constructed with the maximally localized Wannier functions (MLWFs) and surface Green's function methods~\cite{Qui-iter-sche,SurfaceGF,wu2017wanniertools}, we have calculated the corresponding surface states of this system with SOC to identify its topological properties (see Section~6 of the Appendix).
As shown in Figs.~\ref{fig:surface}(a) and \ref{fig:surface}(b), the 2D Dirac cone is obtained on (010) and (001) surfaces. Thus, the calculated surface states agree well with the analysis of the symmetry indicators, which confirm the STI nature of NaCaBi.

\begin{figure}[th]
\begin{center}
\includegraphics[width=0.45\textwidth]{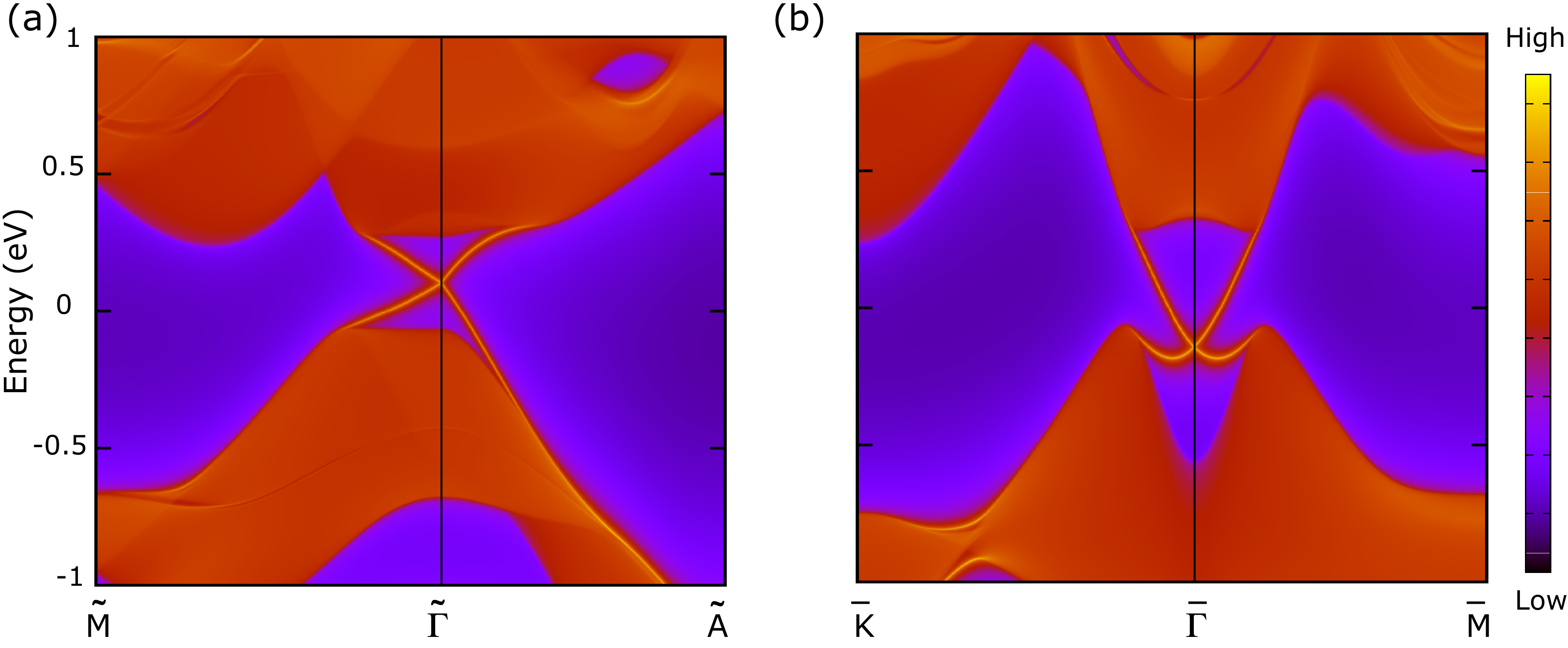}
\caption{%
(a,b) Surface states of NaCaBi terminated in the (010) surface and (001) surface, respectively.
Both (010) and (001) surfaces exhibit a 2D Dirac cone.
}
\label{fig:surface}
\end{center}
\end{figure}

\subsection{ Low-energy effective model }
We find that the topologically nontrivial nature is determined by the low-energy dispersions near $E_F$ around $\Gamma$, which motivates us to build the effective $\mathbf{k}\cdot\mathbf{p}$ Hamiltonian to get further insights in this system. In the presence of SOC, the irreps for the lower energy bands at $\Gamma$ are $\Gamma_{7}^{+}$ and $\Gamma_{7}^{-}$, respectively
(labeled in the double point group of $D_{6h}$).

Thus, we can construct the $4\times 4$ Hamiltonian under the bases of the two irreps (\ie $|+,\uparrow\rangle;|+,\downarrow\rangle;|-,\uparrow\rangle$ and $|-,\uparrow\rangle$) with the theory of invariants~\cite{invariant-theory2003}.
Here the $\pm$ denote the parities of the eigenstates and $\uparrow~(\downarrow)$ denotes $j_{z}=+\frac{1}{2}~(-\frac{1}{2})$.
Taking time reversal symmetry and the symmetries of double group $D_{6h}$ into consideration (see details in Section~7 of the Appendix), we derive the following $\bk\cdot\bp$ invariant Hamiltonian in the vicinity of $\Gamma$ (up to the second order of $\vec k$),
\begin{equation}\label{eq:kp}
\begin{split}
H_{0}(\vec{k})&=\epsilon_{0}(\vec{k}){\mathbb I}_4+
\left(
  \begin{array}{cccc}   %
  M(\vec{k})     &               0               &                A_{1}k_{z}        &           A_{2}k_{-}                \\  %
      0          &          M(\vec{k})           &                A_{2}k_{+}        &          -A_{1}k_{z}         \\  %
  A_{1}k_{z}     &          A_{2}k_{-}           &               -M(\vec{k})        &             0                \\  %
  A_{2}k_{+}     &         -A_{1}k_{z}           &                  0               &          -M(\vec{k})         \\  %
  \end{array}
\right)
\end{split}
\end{equation}
where $k_{\pm}=k_{x}\pm ik_{y}$, $\epsilon_{0}(\vec{k})=C+D_{1}k_{z}^{2}+D_{2}k_{\bot}^{2}$ and $M(\vec{k})=M-B_{1}k_{z}^{2}-B_{2}k_{\bot}^{2}$ with $k_{\bot}^{2}=k_{x}^{2}+k_{y}^{2}$.
This is nothing but the well-known 3D TI Hamiltonian in the Bi$_2$Se$_3$ family~\cite{Bi2X3-zhj2009} in the case of $M\cdot B_{1,2}>0$,
the band inversion happens between the two different-parity bands.
By fitting the energy dispersions, we obtain the parameters: $M$=$-$0.4554 eV, $A_{1}$=4.0 eV{\AA}, $A_{2}$= 1.3 eV{\AA}, $B_{1}$=$-$16.7 eV{\AA}$^{2}$, $B_{2}$=18.6 eV{\AA}$^{2}$, $C$=$-$0.0689 eV, $D_{1}$=$-$6.4 eV{\AA}$^{2}$, $D_{2}$=8.5 eV{\AA}$^{2}$.

As aforementioned, the nontrivial topology of this system can be also protected/indicated by $\bar{6}$ symmetry. If we introduce the additional $\delta_1$ ($\delta_2$) term as shown in Eq.~(\ref{eq:kp-per}), $C_{2x}~(C_{2y})$ and $M_{y}~(M_{x})$ will be broken. Therefore, by simply adding the two additional terms (both of which break $C_{6z}$ and $\cal I$), the $\bar{6}$-invariant Hamiltonian is obtained below,
\begin{equation}\label{eq:kp-per}
\begin{split}
&H(\vec{k})=H_{0}(\vec{k})+H_1(\delta_1+i\delta_2,\vec{k}) \\
&H_1(z,\vec{k})=
\left(
  \begin{array}{cccc}   %
  0 & 0 & 0 & -z^*k_+^2 \\
  0 & 0 & z k_-^2 & 0 \\
  0 & z^* k_+^2 & 0& 0 \\
 -z k_-^2 & 0 & 0& 0 \\
  \end{array}
\right)
\end{split}
\end{equation}
In the presence of $\bar{6}$ symmetry, the $C_{3z}$ and $M_z$ symmetries are still preserved. Thus, one can check that in the case of $M\cdot B_{1,2}>0$, the mirror Chern number of the $k_z=0$ plane is 1.
On the other hand, one can also compute the symmetry indicators of SG 174 ($P\bar 6$), which is founed to be $z_{3m,0}=1$.
This implies that the $\mathbb Z_2$ invariant for the $k_z=0$ plane is 1. Combined with the trivial $\mathbb Z_2$ invariant for the $k_z=\pi$ plane, one can conclude that  $z_{3m,0}=1$ and $z_{3m,\pi}=0$ yield the STI nature of $P6_{3}/mmc$ NaCaBi.
Suppose that we move the two Na atoms at $2a$ Wyckoff positions reversely along $z$ axis a little (\eg two percent of the third lattice vector along $z$ axis), this distortion preserves $\bar{6}$ symmetry, but breaks inversion symmetry and $C_{6z}$ symmetry. In this disturbed structure, the $\delta_1$ and $\delta_2$ are fitted to be 0.4 eV\AA$^2$ and 0 eV\AA$^2$. The STI phase is indicated by the symmetry indicators: $z_{3m,0}=1$ and $z_{3m,\pi}=0$, since there is no band crossing occurring in this process.

\section{DISCUSSION}
\begin{figure*}[!t]
\begin{center}
\includegraphics[width=0.95\textwidth]{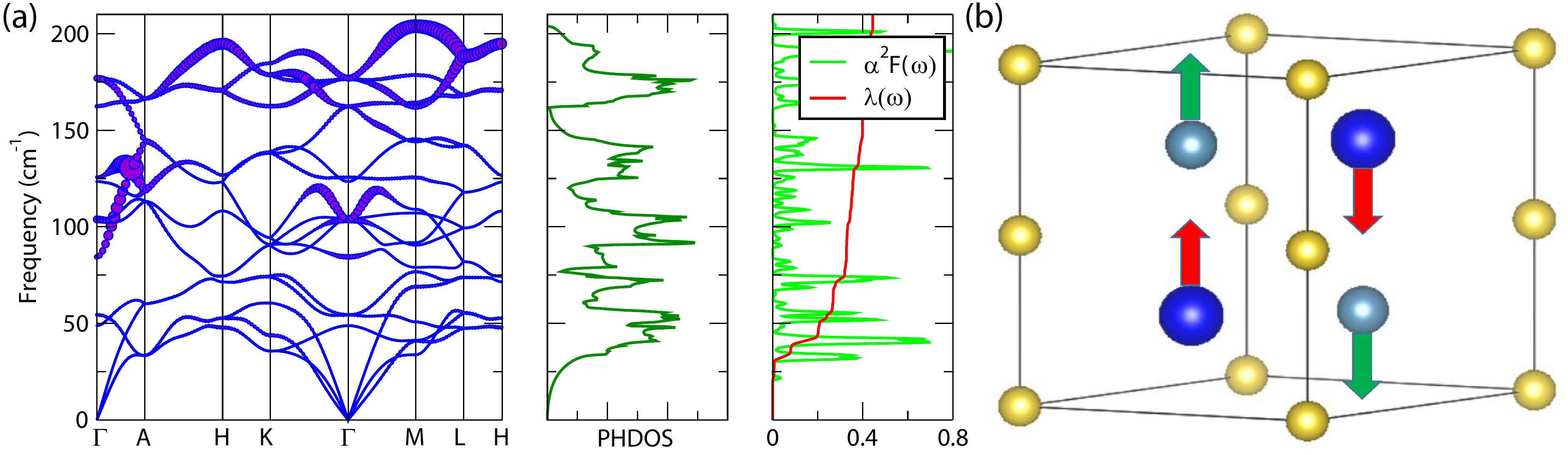}
\caption{%
Electron-phonon coupling calculations of $P6_{3}/mmc$ NaCaBi.
(a) Phonon spectra, phonon density of states (PDOS) and Eliashberg spectral functions $\alpha^{2}F(\omega)$ together with the electron-phonon integral $\lambda(\omega)$ in the left, middle and right panels, respectively.
(b) The $B_{2g}$ vibrational mode.
}
\label{fig:B2g}
\end{center}
\end{figure*}
In carrier-doped TIs, odd-parity pairing and nematic pairing have been theoretically proposed to realize topological superconductivity \cite{Fu2010,Fu2014}. Some supporting evidences in Cu-doped Bi$_2$Se$_3$ are revealed in experiments but the pairing symmetry is still under active debate~\cite{Sasaki2011,Kirzhner2012,Bay2012,Levy2013,Matano2016}. In analogous to Bi$_{2}$Se$_{3}$, superconductivity and topological paring states may be achieved with electron doping in NaCaBi.
One prominent feature of $P6_{3}/mmc$ NaCaBi is that the band inversion is kept with SOC and is significantly large, resulting a double-peak valence band around $\Gamma$, as shown in Fig.~\ref{fig:2}(c). Within a relatively large hole doping region, the Fermi surface is a torus around the $\Gamma$ point rather than spheres in other TIs, as shown in Fig.~\ref{fig:2}(d). As the topology of the Fermi surface is distinct from electron-doped case, the pairing state can be also different and may be unconventional. For example, for the intra-orbital spin singlet pairing $c_{+\uparrow}c_{+\downarrow}-c_{-\uparrow}c_{-\downarrow}$,  which belongs to a trivial irrep, two nodal lines of gap functions in band space are expected on the torus. We leave the question of determining possible pairing states by solving effective models to future work. Nevertheless, we investigate the superconducting properties by electron-phonon coupling calculations and the Allen-Dynes modified McMillian equation~\cite{Tc-Allen}. Considering 0.06 hole dopping in NaCaBi system at ambient pressure, Tc is estimated to be about 1.5 K with a commonly used screened Coulomb potential $\mu^{*} = 0.10$.  Most of the contributions to electron-phonon coupling constant arise from the $B_{2g}$ vibrational mode, \ie vibrations of Ca atoms and Bi atoms along $z$ axis, as shown in Fig.~\ref{fig:B2g}(b).

In summary, we propose a new NaCaBi family which are STIs with large SOC-induced band gaps, among which KBaBi, KSrBi, RbBaBi and RbSrBi are
stable at ambient pressure, both in thermodynamics and lattice dynamics. In terms of the NaCaBi system, the corresponding symmetry analysis indicates that there exists a clear band inversion between two opposite-parity bands near $E_F$.
Without SOC, the band inversion gives rise to a nodal ring around $\Gamma$ in the $k_{z}=0$ plane. Upon including SOC, it becomes a STI with a SOC-induced band gap of 0.34 eV, which is larger than the energy scale of room temperature.
The calculated symmetry indicators of SG 194 indicate that the nontrivial topology of NaCaBi can be revealed by not only the inversion symmetry,  but also $\bar{6}$ symmetry.
As expected, 2D Dirac cones are obtained in the (010)-surface and (001)-surface spectra.
Due to the large SOC-induced band gap in the STI phase, it is promising to expect clear experimental evidences of the topological surface states from angle-resolved photoemission spectroscopy (ARPES) and scanning tunnelling microscopy
in the future. With hole doping, the contour surface of the valence band near $E_{F}$ is a torus, which may bring unconventional superconducting pairing.

\section{ ACKNOWLEDGMENTS }
We thank the fruitful discussions with Lu Liu, Tong Chen and Qinyan Gu.
This work is supported by the National Natural Science Foundation of China (Grants Nos. 11974395, 11974162, 11925408, 11834006 and 11921004).  J.S. also gratefully acknowledges financial support from the National Key R\&D Program of China (Grant Nos. 2016YFA0300404) and the Fundamental Research Funds for the Central Universities. Part of the calculations were carried out using supercomputers at the High Performance Computing Center of Collaborative Innovation Center of Advanced Microstructures, the high-performance supercomputing center of Nanjing University. Z.W. and H.W. also acknowledge support from the National Key Research and Development Program of China (Grant Nos. 2016YFA0300600, 2016YFA0302400, and 2018YFA0305700), the K. C. Wong Education Foundation (GJTD-2018-01), the Strategic Priority Research Program of Chinese Academy of Sciences (Grant No. XDB33000000), and the center for Materials Genome.

\section*{APPENDIX}
\subsection*{1. Calculation methods}
\label{sup:A}

\textbf{\emph{Ab initio}} \textbf{calculations.} We performed first-principles calculations based on the density functional theory (DFT) using projector augmented wave (PAW) method implemented in the Vienna ab initio simulation package (VASP)~\cite{kresse1999} for the calculations of the structure optimization and electronic structures. The generalized gradient approximation (GGA), as implemented in the Perdew-Burke-Ernzerhof (PBE) functional~\cite{GGA-PBE1996} was adopted. The cutoff parameter for the wave functions was 850 eV. The BZ was sampled by Monkhorst-Pack method~\cite{Monkhorst-BZ1976} with a k-spacing of $0.025\times 2\pi$ \AA$^{-1}$ for the 3D periodic boundary conditions. The phonon calculations were performed by finite displacement method implemented in the PHONOPY code~\cite{phonopy}, with a $2\times2\times2$ supercell. A kmesh of $20\times 20\times 16$ was used to calculate the Fermi surface of the valence band near the Fermi level.  Electron-phonon coupling calculations are performed in the framework of Density functional perturbation theory, as implemented in the quantum-espresso code~\cite{QE2009}.

\begin{figure*}[!htp]
\begin{center}
\includegraphics[width=0.9\textwidth]{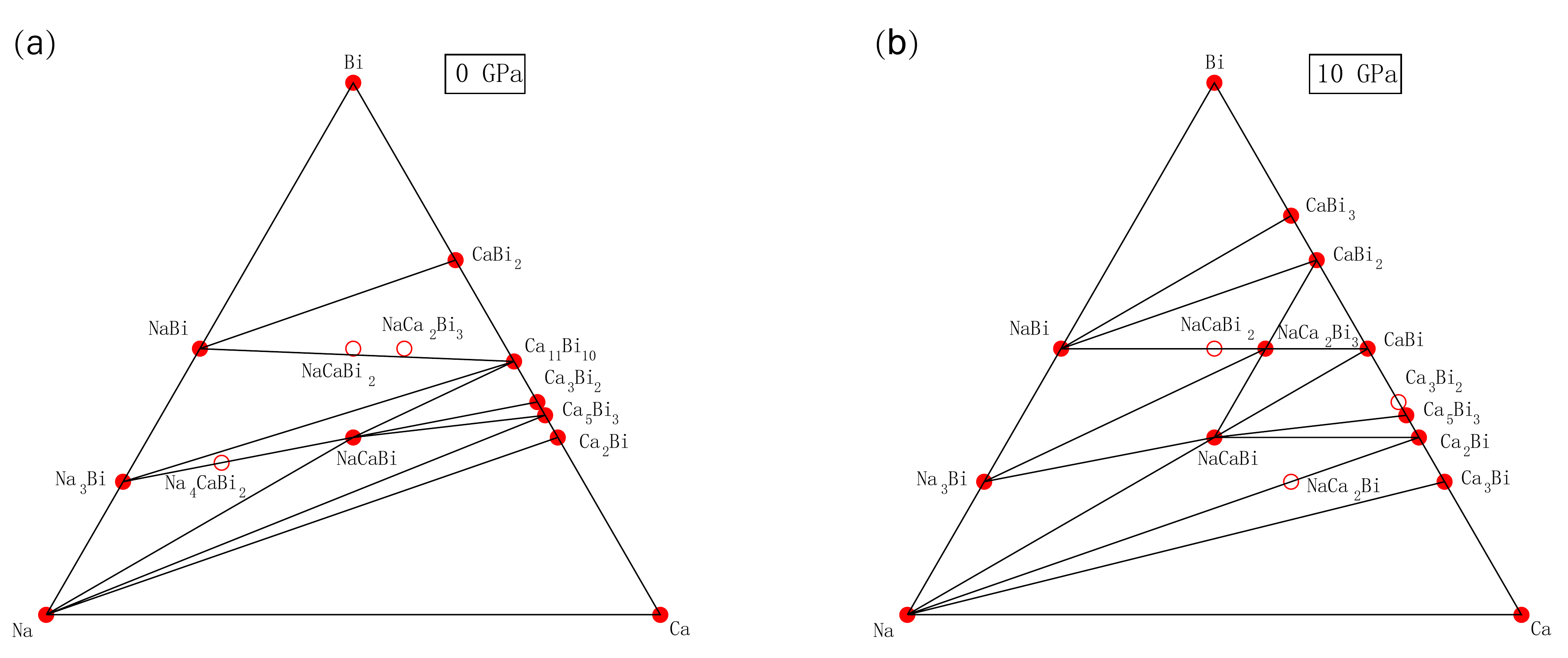}
\caption{%
Phase diagram based on convex hull analysis of formation enthalpies for the Na-Ca-Bi ternary system at (a) 0 GPa and (b) 10 GPa. The solid red circles in the figures denote the stable structures, while the hollow red circles denote the corresponding structures lying above the convex hull which indicates these structures are metastable candidates. The solid red circle in the center of the equilateral triangle in figure (a) and (b) denote the $Pnma$ phase at 0 GPa and $P6_{3}/mmc$ phase at 10 GPa.
}
\label{fig:convex}
\end{center}
\end{figure*}

\begin{table*}[htb]
\centering
\caption{Crystal structures of NaCaBi after structure optimization.}
   \begin{tabular}{c c c c}%
       \hline\hline
          \textbf{SG}       & \textbf{\{$\alpha,\beta,\gamma$\}}                 & \textbf{\{a,b,c\}}                               & \textbf{Wyckoff positions}   \\
\hline
            $P6_{3}/mmc$    &  $\alpha=\beta=90{}^{\circ},\gamma=120{}^{\circ}$  & $\{a=b=5.566{\AA},c=6.804{\AA}\}$                & \{$2a,2c,2d$\} \\
            $P3m1$         &  $\alpha=\beta=90{}^{\circ},\gamma=120{}^{\circ}$  & $\{a=b=5.3431{\AA},c=7.41{\AA}\}$                & \{$1a\oplus1c,1b\oplus1c,1a\oplus1b$\} \\
            $Pnma$          &  $\alpha=\beta=\gamma=90{}^{\circ}$                & $\{a=8.088{\AA},b=4.784{\AA},c=9.681{\AA}\}$     & \{$4c,4c,4c$\} \\
            $F\bar{4}3m$   &  $\alpha=\beta=\gamma=90{}^{\circ}$                & $\{a=b=c=7.530{\AA}\}$                           & \{$4a,4b,4c$\} \\
            $P\bar{4}2m$    &  $\alpha=\beta=\gamma=90{}^{\circ}$                & $\{a=b=7.5706{\AA},c=7.5032{\AA}\}$              & \{$1a\oplus1b\oplus2f,1c\oplus1d\oplus2e,4n$\} \\
       \hline\hline
  \end{tabular}
  \begin{tablenotes}
        \footnotesize
        \item[1] 1. The ``SG'' column denotes the space group in which NaCaBi crystallises into.
        \item[2] 2. \{$\alpha,\beta,\gamma$\} and \{a,b,c\} denote the lattice parameters of the conventional cell (while not primitive cell) in each phase.
        \item[3] 3. The ``Wyckoff positions'' column gives the Wyckoff positions of Na, Ca and Bi in each phase. ``$1a\oplus1c$'', ``$1b\oplus1c$'' and ``$1a\oplus1b$'' in $P3m1$ phase indicates that there exist two nonequivalent ```Na''/```Ca''/```Bi'' atoms occupying (1a and 1c)/(1b and 1c)/(1a and 1b) Wyckoff positions in this phase, respectively.
      \end{tablenotes}
\label{table:str}
\end{table*}

\subsection*{2. Crystal structure searching}
\label{sup:B}
We perform a variable composition calculation to get the phase diagrams of Na-Ca-Bi ternary system at ambient condition (0 GPa) and 10 GPa. At ambient condition, we find $Pnma$ NaCaBi is the unique ternary phase lying in the convex hull (denoted by the solid red circle in the center of the equilateral triangle), as shown in Fig.~\ref{fig:convex}(a). Meanwhile, there exist three other metastable ternary candidates, i.e., P4/mmm-NaCaBi$_2$, Cmmm NaCa$_{2}$Bi$_{3}$ and R3m Na$_{4}$CaBi$_{2}$, above the convex hull (denoted by the hollow red circles in the the equilateral triangle). The corresponding enthalpies of these metastable ternary candidates above convex hull (denoted by $\triangle H$) of $P4/mmm$ NaCaBi$_{2}$, $Cmmm$ NaCa$_{2}$Bi$_3$ and $R3m$ Na$_{4}$CaBi$_2$ are 16.0 meV/atom, 13.3 meV/atom and 19.7 meV/atom, respectively. We have listed the structure parameters of these ternary structures in Table.~\ref{table:str-0gpa}.

At 10 GPa, we find $P6_{3}/mmc$ NaCaBi and $Cmmm$ NaCa$_{2}$Bi$_{3}$ (denoted by two solid red circles lying inside the equilateral triangle) are two stable ternary structures lying in the convex hull, as shown in Fig.~\ref{fig:convex}(b). In addition, the convex hull analysis indicates that there are three other structures ($C2/m$ Ca$_{3}$Bi$_{2}$, $Cmmm$ NaCaBi$_{2}$ and $P4/mmm$ NaCa$_{2}$Bi) lying above the convex hull, which indicates they are metastable candidates. The corresponding structure parameters are listed in Table.~\ref{table:str-10gpa}. It should be noted that $C2/m$ Ca$_{3}$Bi$_{2}$ and $Cmmm$ NaCaBi$_{2}$ are metastable candidates with only 0.64 meV/atom and 1.2 meV/atom above the convex hull. We also remind that 
binary Na-Bi and binary Ca-Bi phases in the convex hull are almost in accordance with earlier works~\cite{Ca-Bi-2015,NaBi-system-2015}.

\begin{table*}[!htp]
\centering
\caption{Crystal structures of Na-Ca-Bi ternary system near the convex hull at ambient condition.}
\begin{tabular}{c c c c c}%
\hline\hline
\textbf{SG}       & \textbf{\{$\alpha,\beta,\gamma$\}}                 & \textbf{\{a,b,c\}}                          & \textbf{Wyckoff positions}  & \textbf{$\triangle H$} \\
\hline
$Pnma$-NaCaBi          &  $\alpha=\beta=\gamma=90{}^{\circ}$                & $\{a=8.088{\AA},b=4.784{\AA},c=9.681{\AA}\}$     & \{$4c,4c,4c$\}  &   0 \\
$P4/mmm$-NaCaBi$_{2}$  &  $\alpha=\beta=\gamma=90{}^{\circ}$  & $\{a=b=5.2019{\AA},c=9.6268{\AA}\}$                & \{$1c,1a,2e$\} &   16 \\
$Cmmm$-NaCa$_{2}$Bi$_{3}$  &  $\alpha=\beta=\gamma=90{}^{\circ}$  & $\{a=15.2393{\AA},b=5.1890{\AA},c=4.5445{\AA}\}$                & \{$2a,4g,4h\oplus2c$\} &   13.3\\
$R3m$-Na$_{4}$CaBi$_{2}$  &  $\alpha=\beta=90{}^{\circ},\gamma=120{}^{\circ}$  & $\{a=b=5.4382{\AA},c=26.6190{\AA}\}$  & \{$3a\oplus3a\oplus3a\oplus3a,3a,3a\oplus3a$\} &   19.7\\
\hline\hline
\end{tabular}
  \begin{tablenotes}
        \footnotesize
        \item[1] 1. $\triangle H$ column is under the unit of meV.
      \end{tablenotes}
\label{table:str-0gpa}
\end{table*}

\begin{table*}[!htp]
\centering
\caption{Crystal structures of Na-Ca-Bi ternary system near the convex hull at 10 GPa.}
\begin{tabular}{c c c c c}%
\hline\hline
\textbf{SG}       & \textbf{\{$\alpha,\beta,\gamma$\}}                 & \textbf{\{a,b,c\}}                          & \textbf{Wyckoff positions}  & \textbf{$\triangle H$} \\
\hline
$P6_{3}/mmc$-NaCaBi    &  $\alpha=\beta=90{}^{\circ},\gamma=120{}^{\circ}$  & $\{a=b=5.1857{\AA},c=6.2652{\AA}\}$                & \{$2a,2c,2d$\}  &   0 \\
$Cmmm$-NaCa$_{2}$Bi$_{2}$  &  $\alpha=\beta=\gamma=90{}^{\circ}$  & $\{a=14.6714{\AA},b=4.7828{\AA},c=4.3440{\AA}\}$           & \{$2a,4g,4h\oplus2c$\} &   0\\
$Cmmm$-NaCaBi$_{2}$   &  $\alpha=\beta=\gamma=90{}^{\circ}$  & $\{a=6.7302{\AA},b=8.7712{\AA},c=3.3638{\AA}\}$           & \{$2c,2d,4e$\} &   1.2\\
$P4/mmm$-NaCa$_{2}$Bi &  $\alpha=\beta=\gamma=90{}^{\circ}$  & $\{a=b=4.7903{\AA},c=4.3642{\AA}\}$                & \{$2c,2d,4e$\} &   9.5 \\
\hline\hline
\end{tabular}
\label{table:str-10gpa}
\end{table*}

\begin{figure*}[!htp]
\begin{center}
\includegraphics[width=0.325\textwidth]{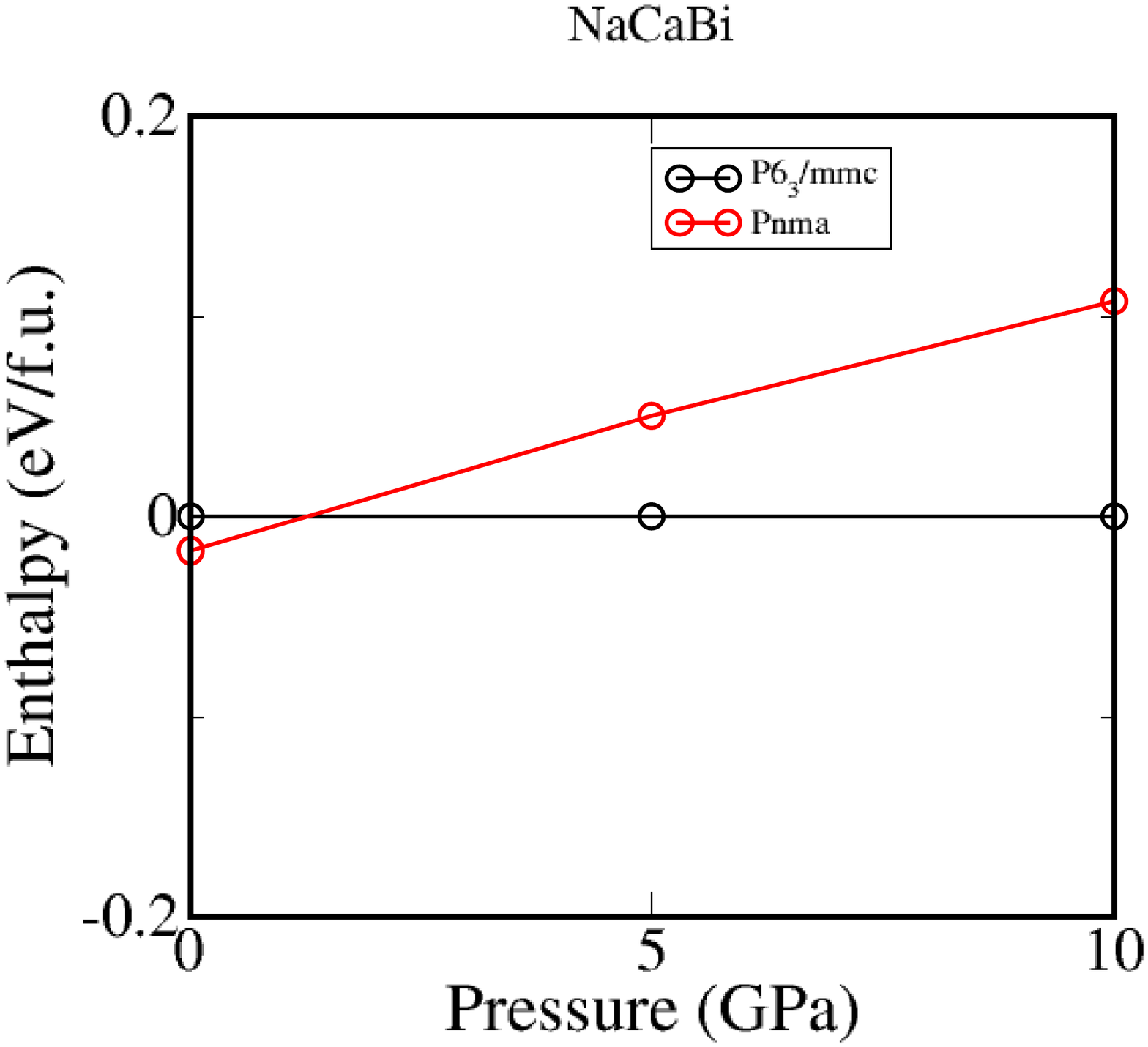}
\includegraphics[width=0.325\textwidth]{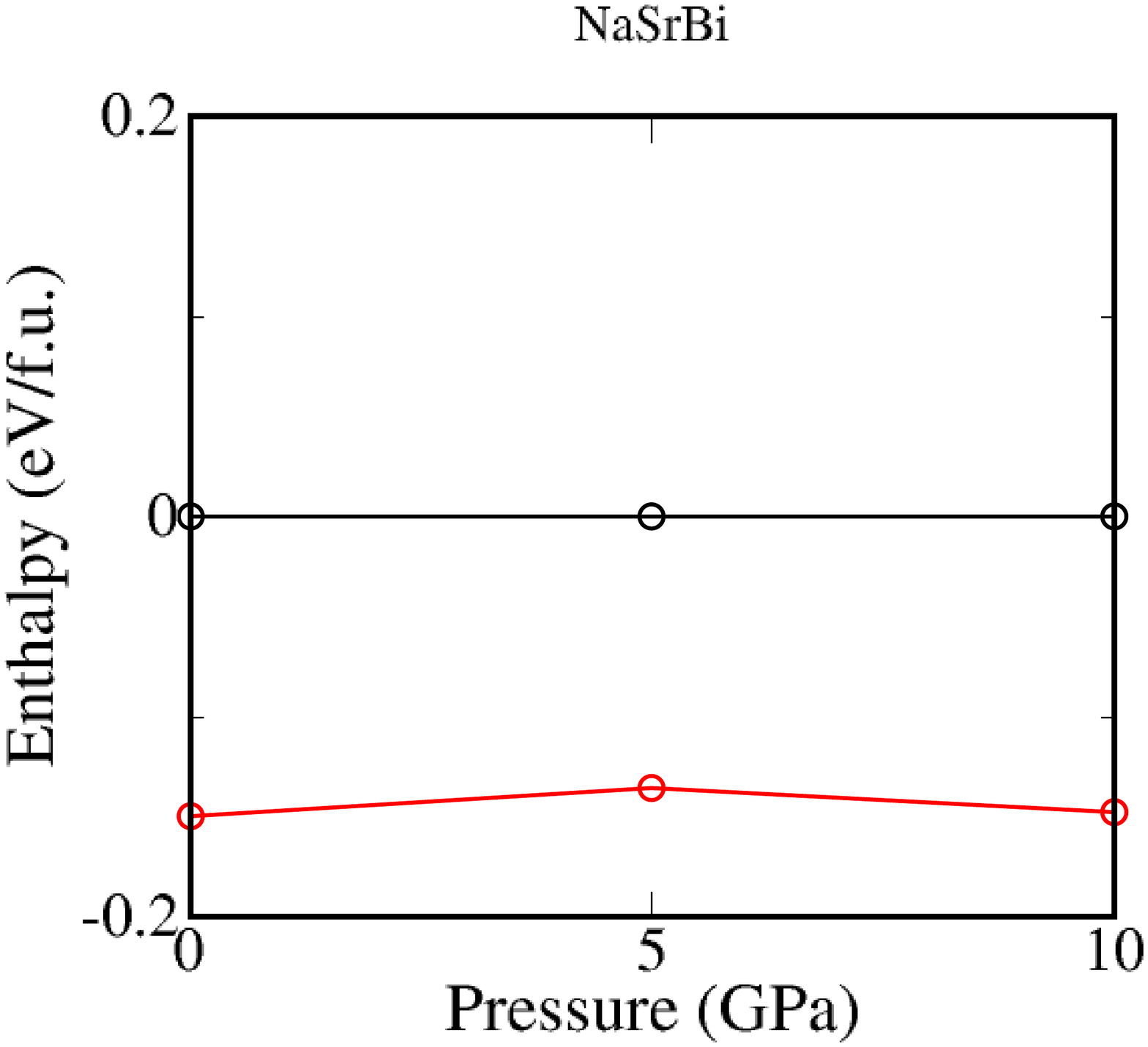}
\includegraphics[width=0.325\textwidth]{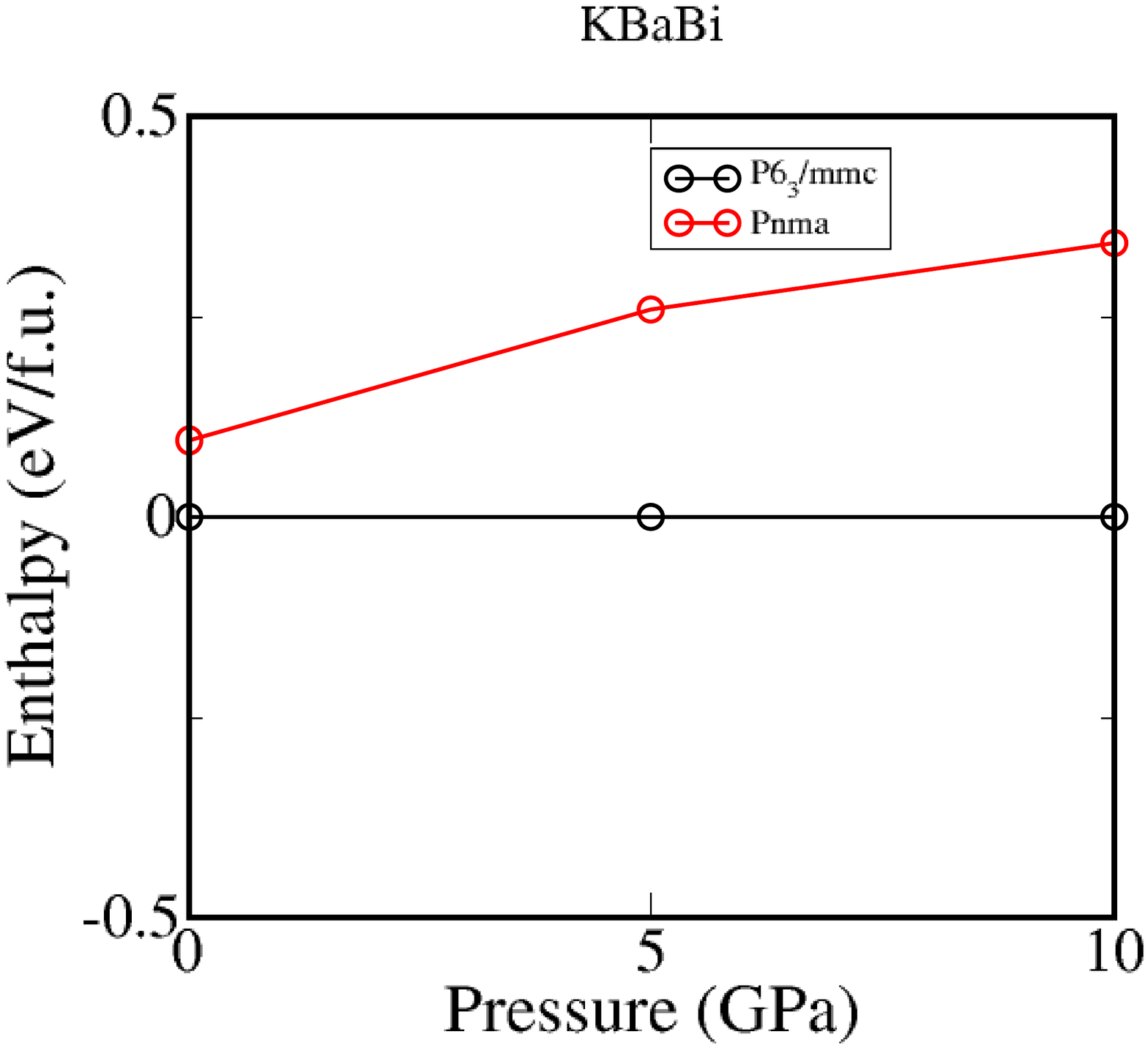}
\includegraphics[width=0.325\textwidth]{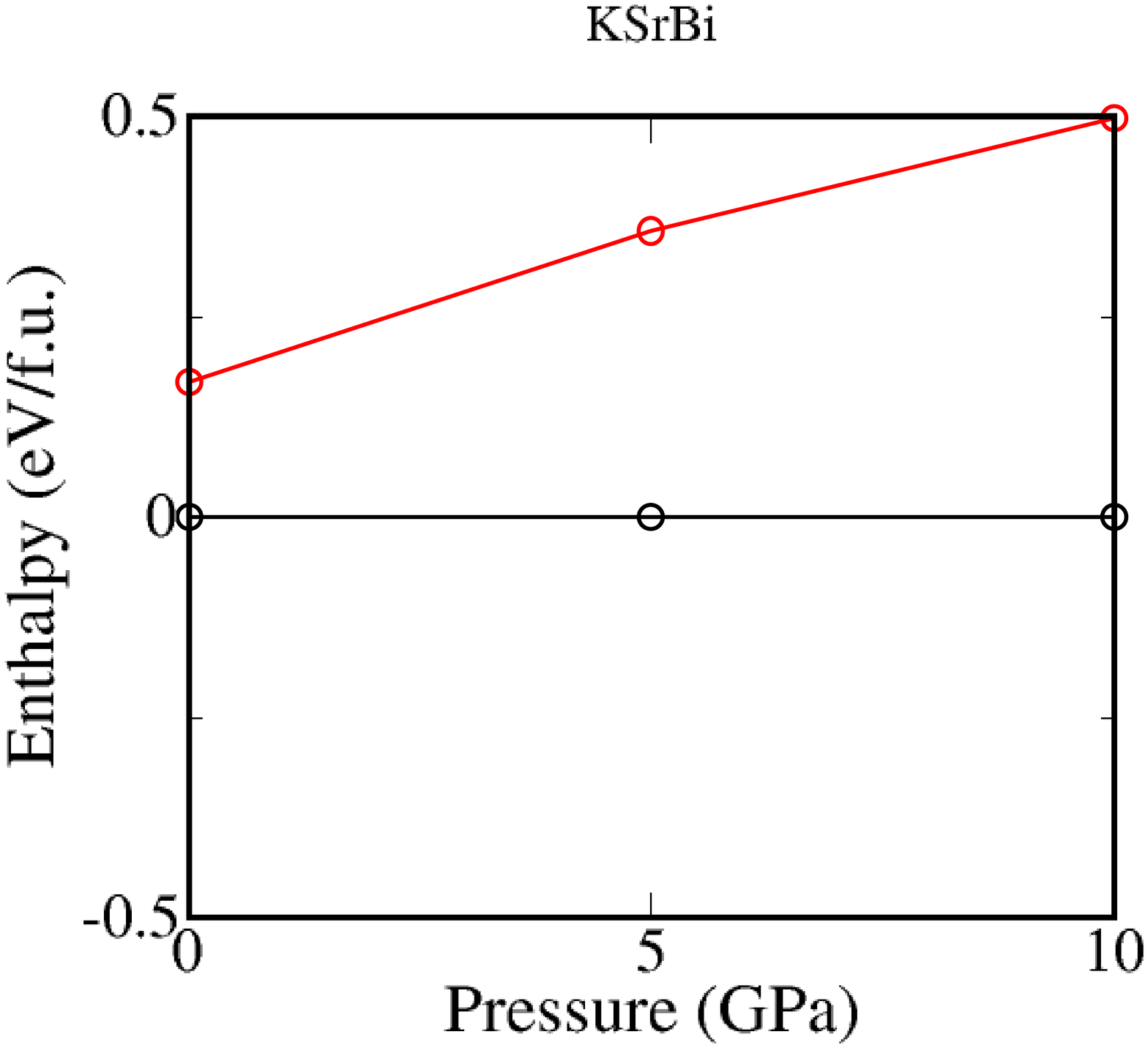}
\includegraphics[width=0.325\textwidth]{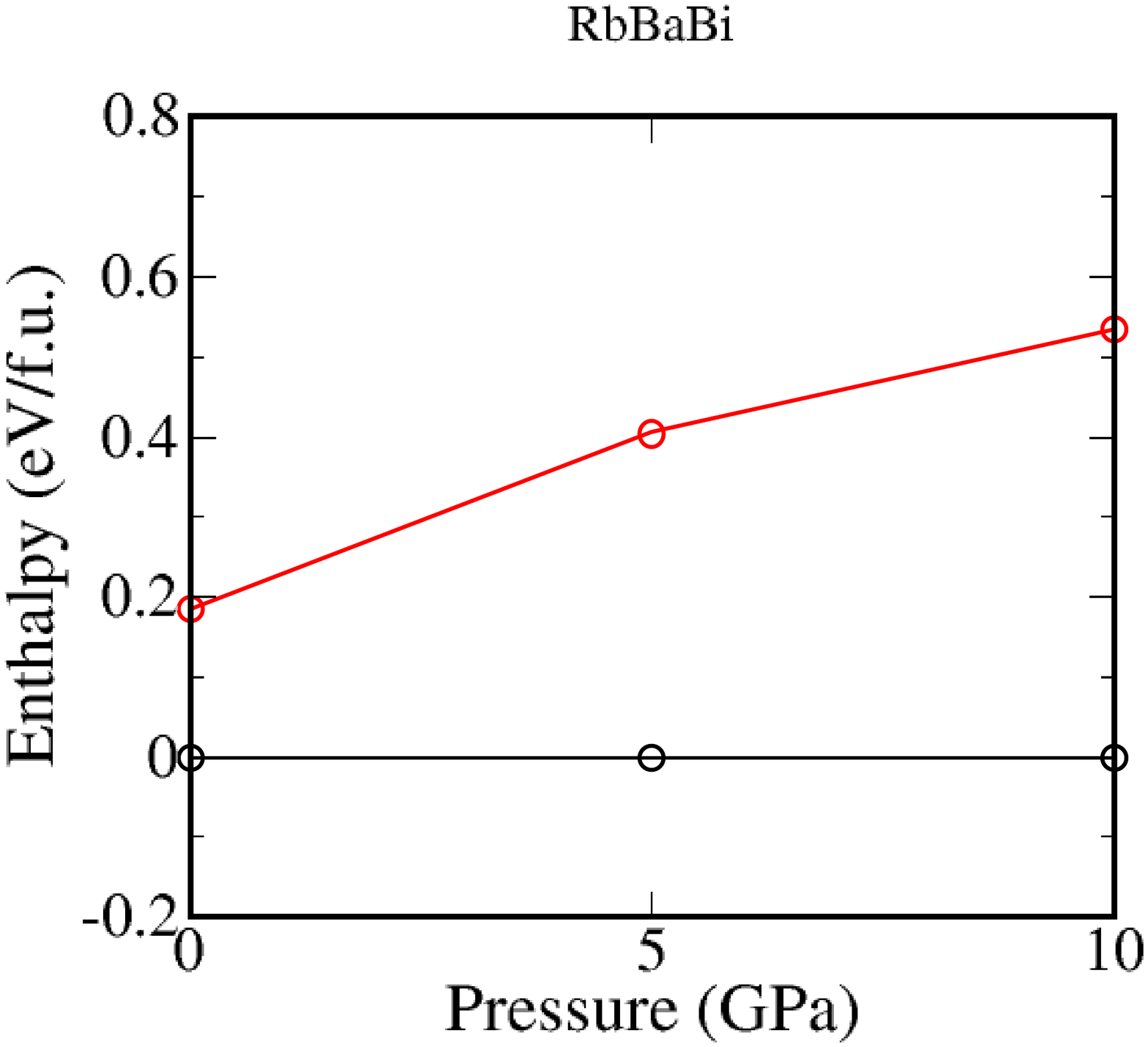}
\includegraphics[width=0.325\textwidth]{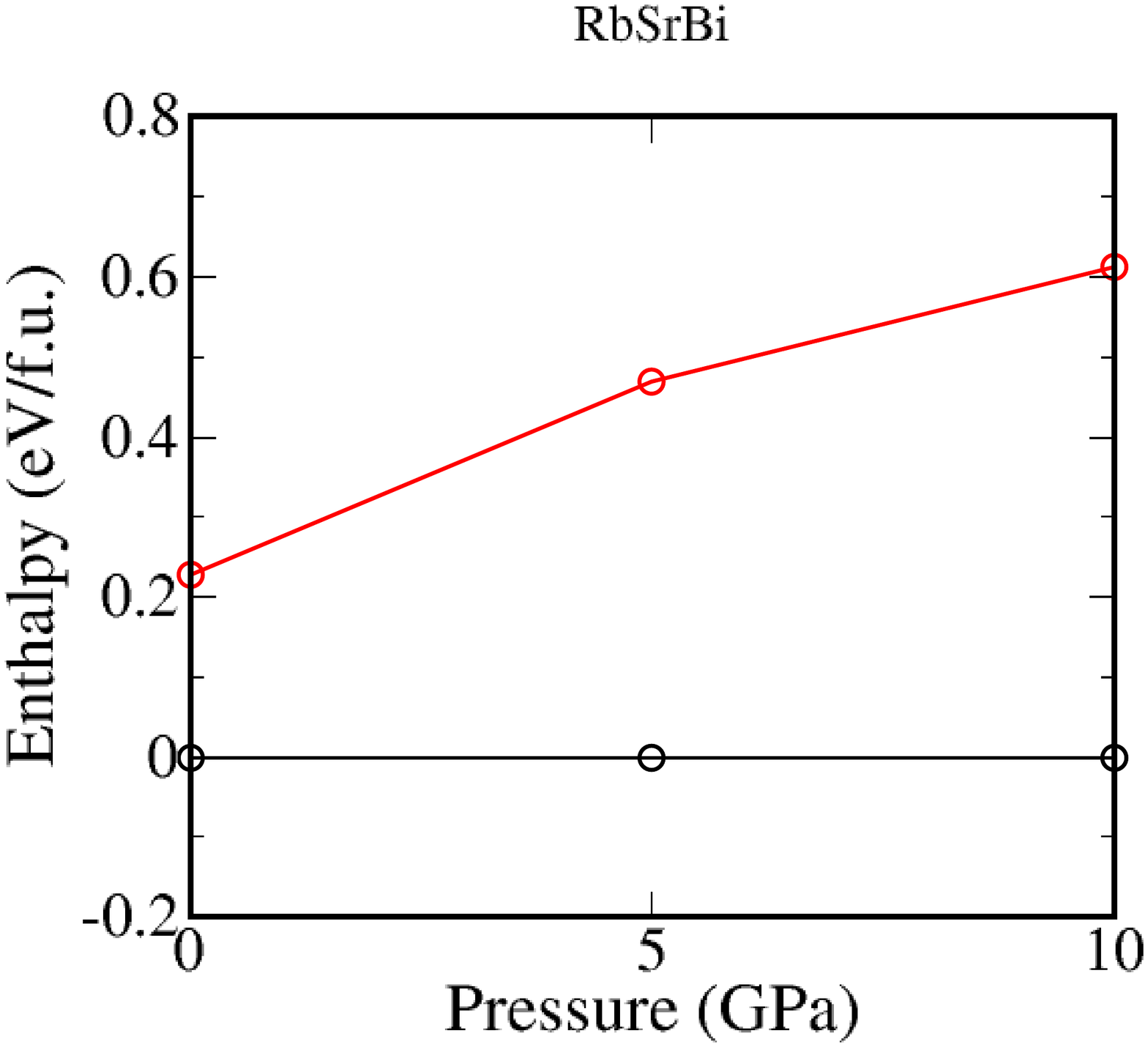}
\caption{%
 Calculated enthalpy between the $Pnma$ phase and $P6_{3}/mmc$ phase vs. pressure for NaCaBi, NaSrBi, KBaBi, KSrBi, RbBaBi and RbSrBi, respectively.
}
\label{fig:pnma}
\end{center}
\end{figure*}
We do not perform ternary structural searching for the other four XYBi candidates, because it will be time consuming. But, similar with NaCaBi, the thermal stability between the $P6_{3}/mmc$ phase and the $Pnma$ phase are discussed. As shown in Fig.~\ref{fig:pnma}, the $Pnma$ phase of NaSrBi is always preferred under pressures from 0 GPa to 10 GPa, while for cases of KBaBi, KSrBi, RbBaBi and RbSrBi, the $P6_{3}/mmc$ phase are always preferred.
It means the STI phase can be obtained in the KBaBi, KSrBi, RbBaBi and RbSrBi systems without any quenched progress.

\subsection*{3. Molecular dynamics}
\label{sup:C}
\begin{figure*}[!htp]
\begin{center}
\includegraphics[width=0.925\textwidth]{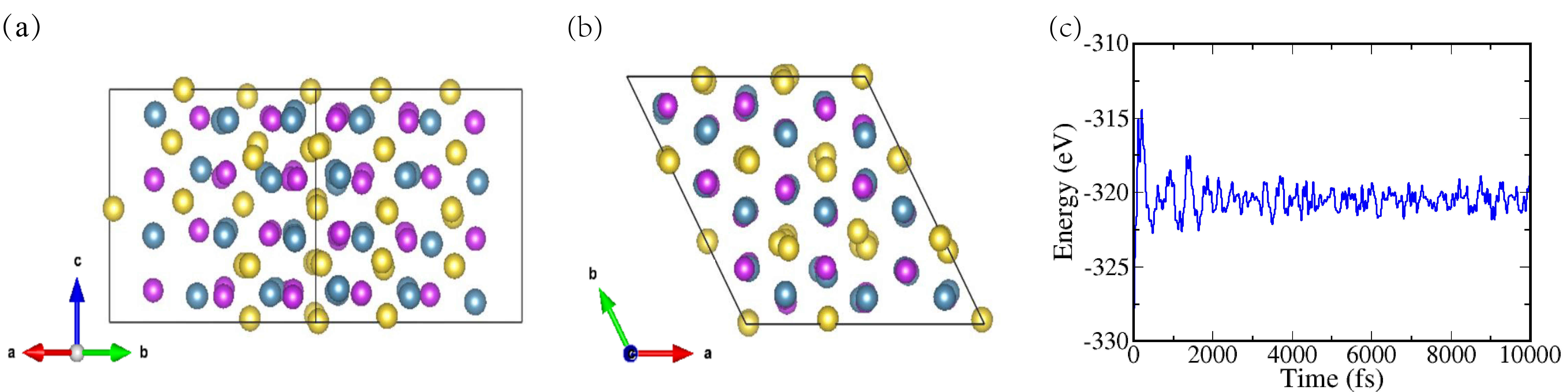}
\caption{%
After 10 ps AIMD simulation, the side view (a) along (001) direction and (b) along (110) direction of the $3 \times 3 \times 2$ supercell of NaCaBi at 600 K. (c) The NVT-molecular dynamic (MD) simulations of the NaCaBi at static pressure and T = 600 K.
}
\label{fig:md}
\end{center}
\end{figure*}
To check the thermal stability of $P6_{3}/mmc$ NaCaBi, we have performed ab initio molecular dynamic (AIMD) simulations with a $3 \times 3 \times 2$ supercell at ambient pressure and T = 600 K. As shown in Fig.~\ref{fig:md}(c), no structural collapse was observed after 10 ps (10000 steps), which indicates the thermal stability of NaCaBi at ambient pressure.

\subsection*{4. Phonon spectra and band structures of XYBi compounds with $P6_{3}/mmc$ structure under low pressures}
\label{sup:D}
\begin{figure*}[!bp]
\begin{center}
\includegraphics[width=0.44\textwidth]{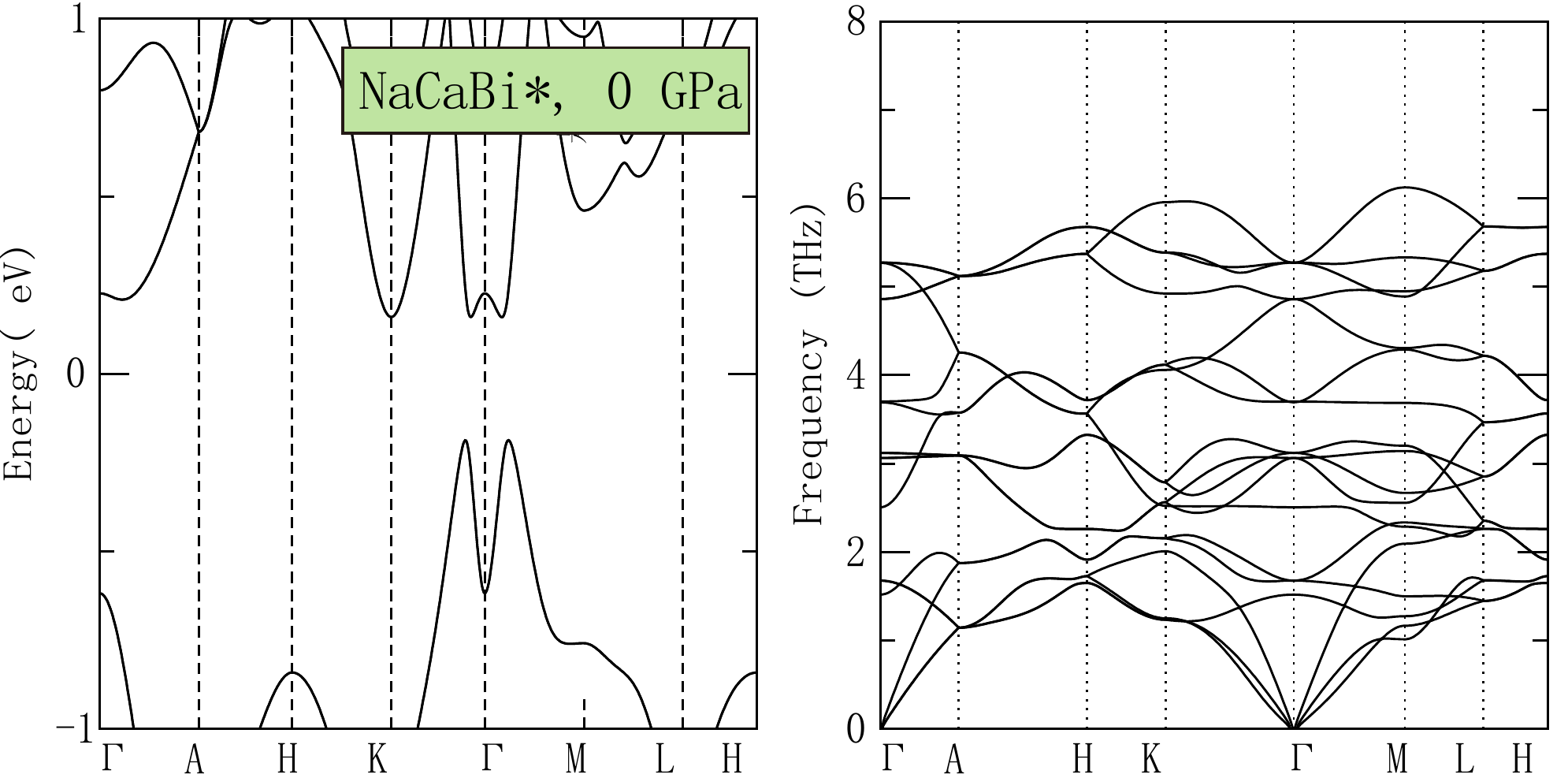}
\includegraphics[width=0.44\textwidth]{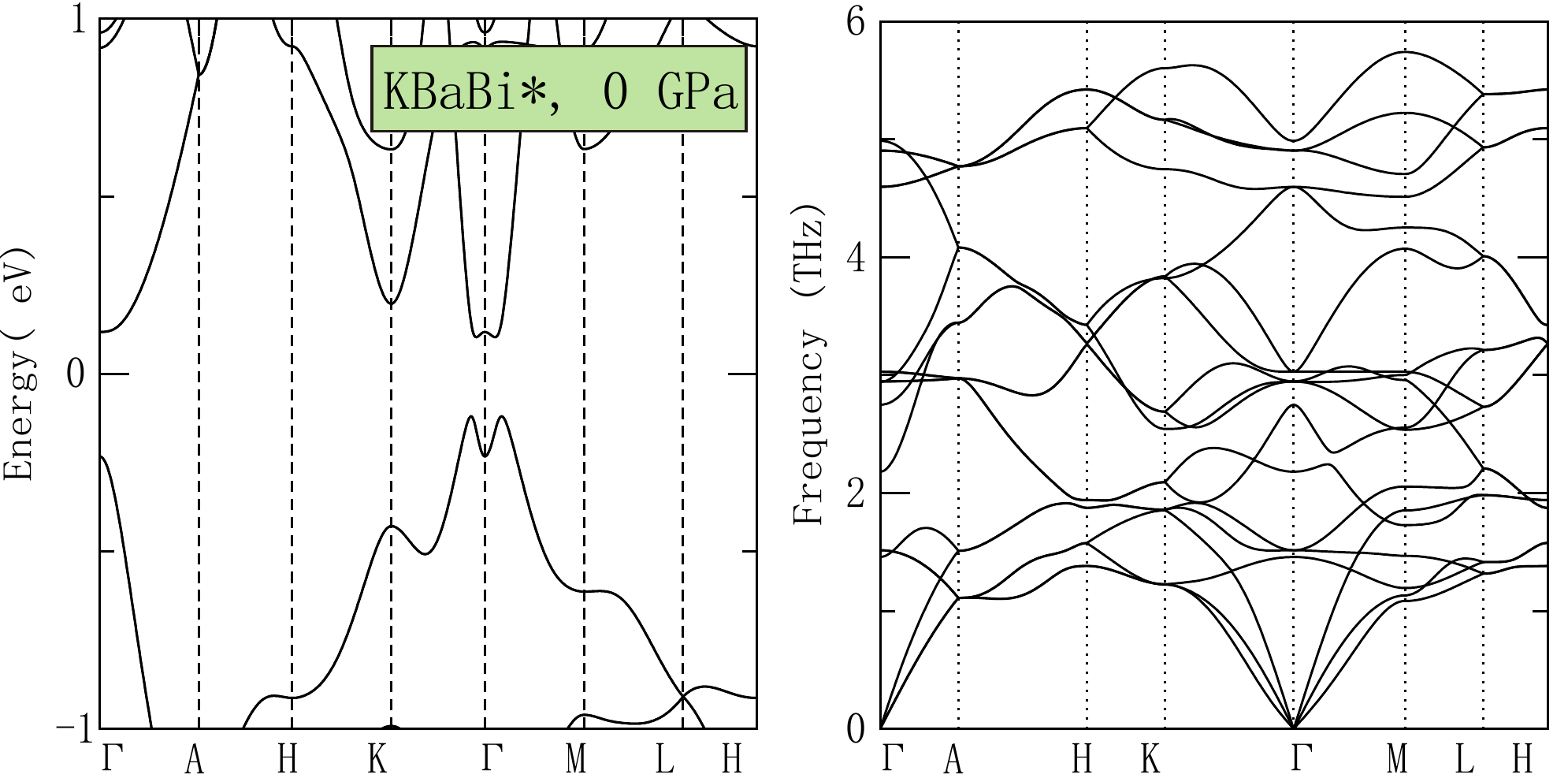}
\includegraphics[width=0.44\textwidth]{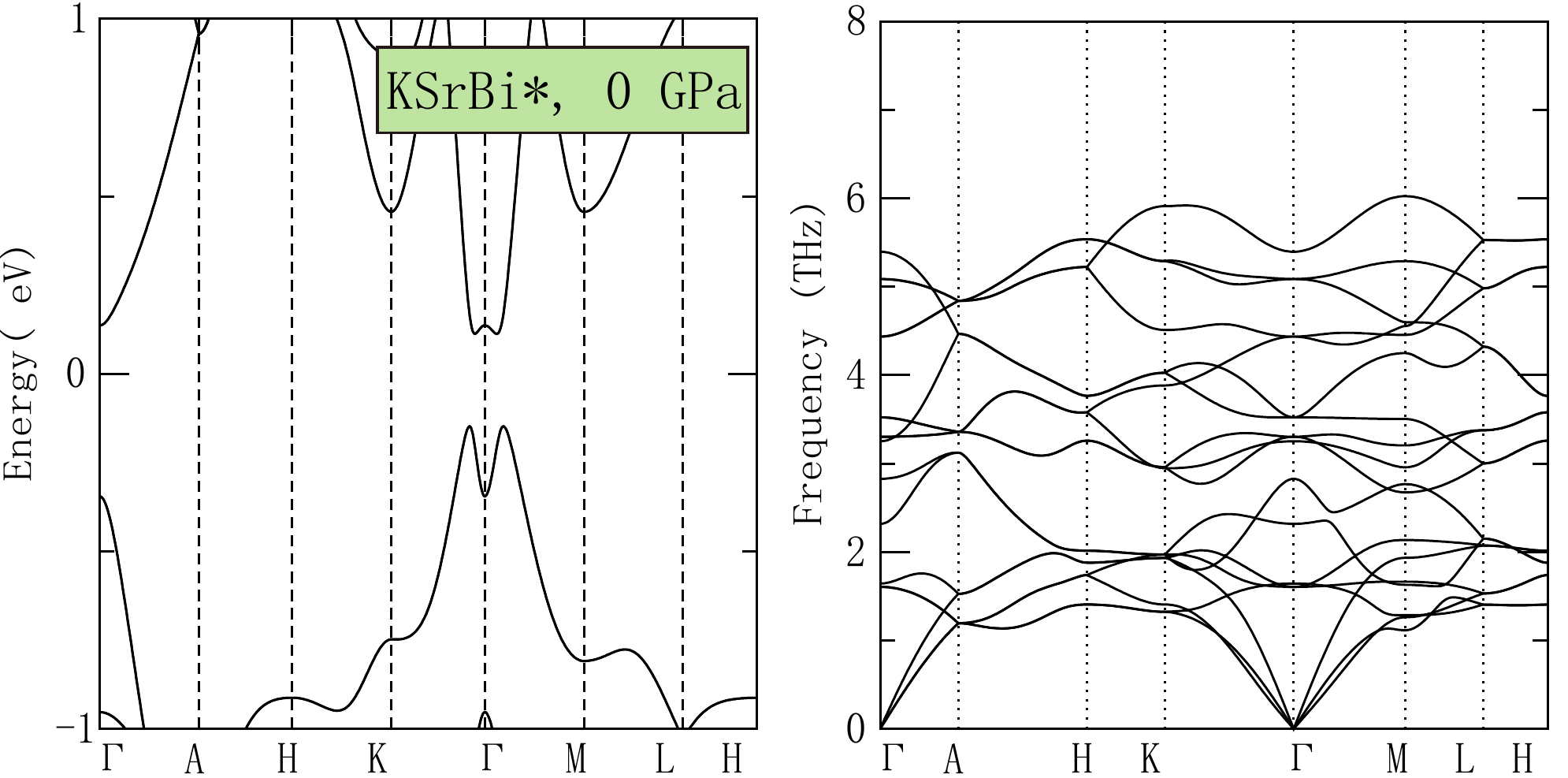}
\includegraphics[width=0.44\textwidth]{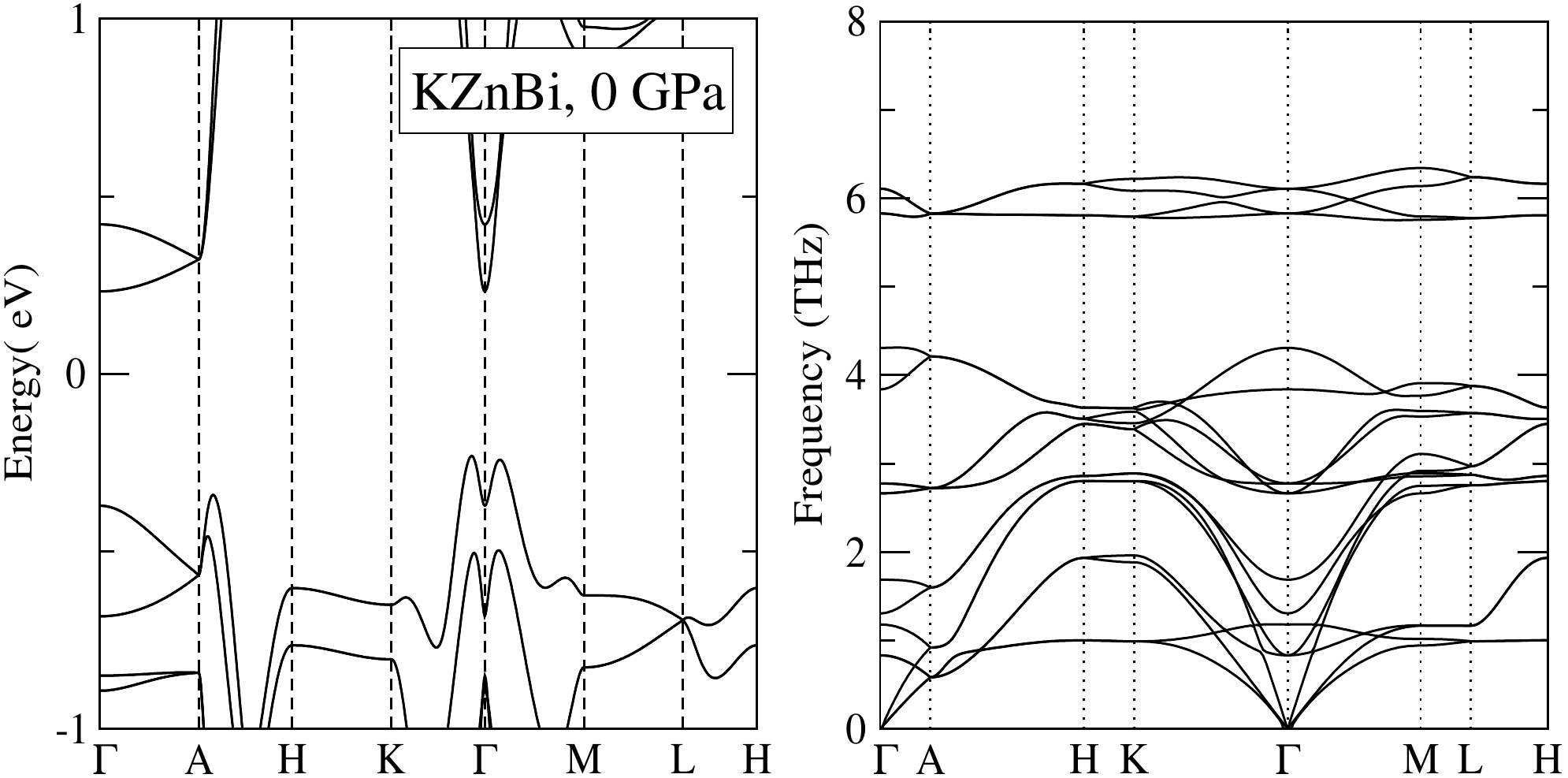}
\includegraphics[width=0.44\textwidth]{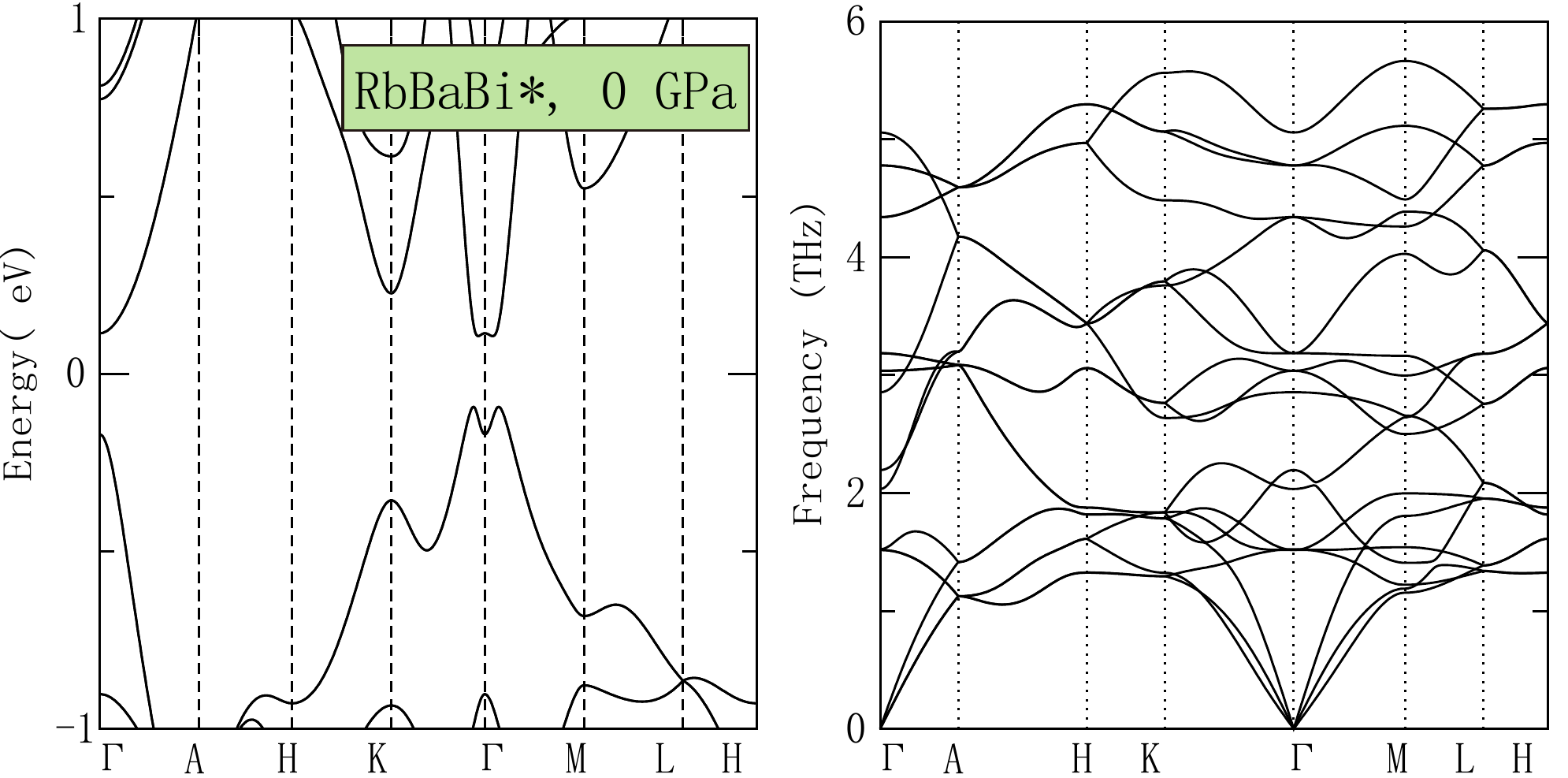}
\includegraphics[width=0.44\textwidth]{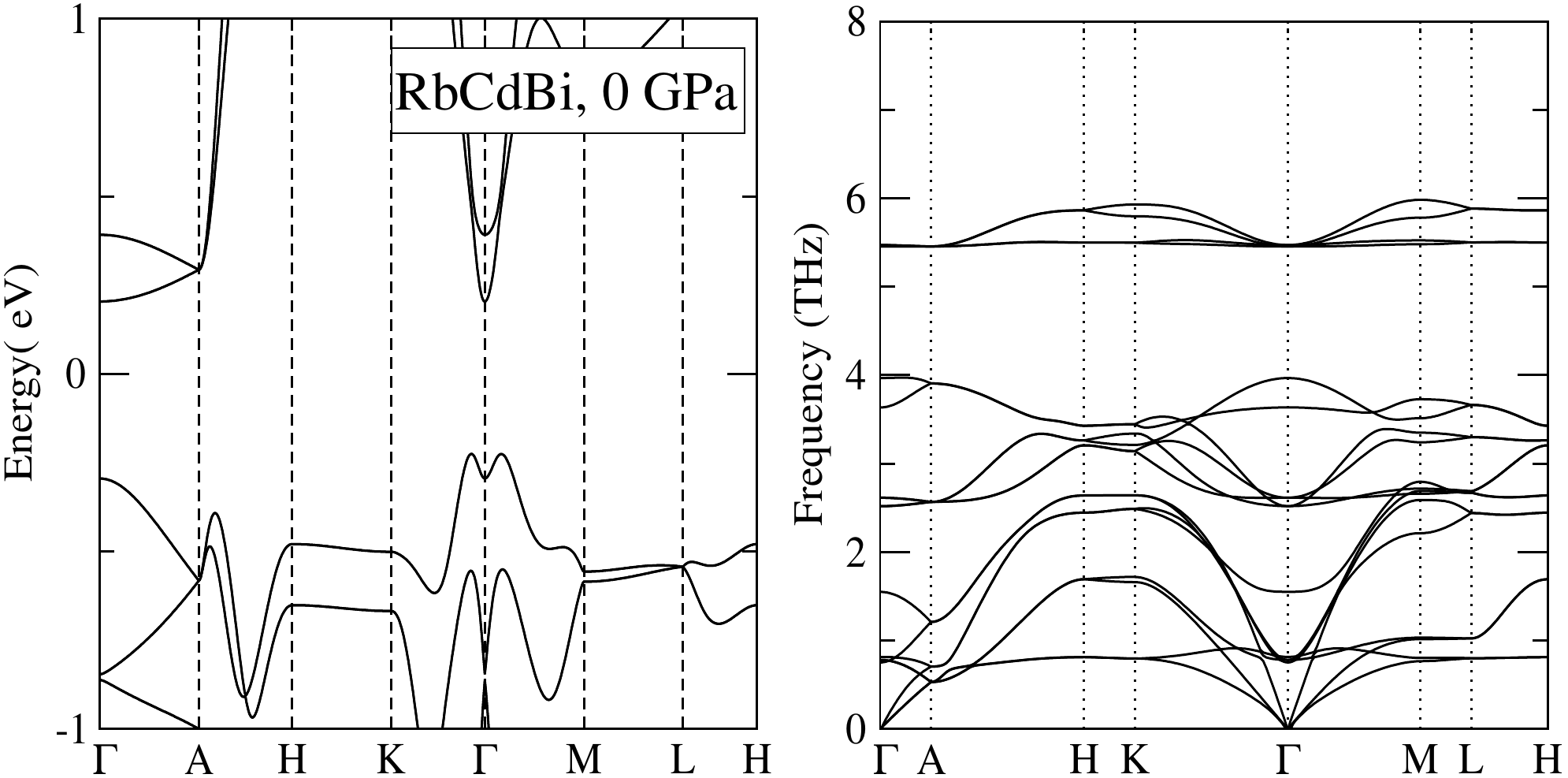}
\includegraphics[width=0.44\textwidth]{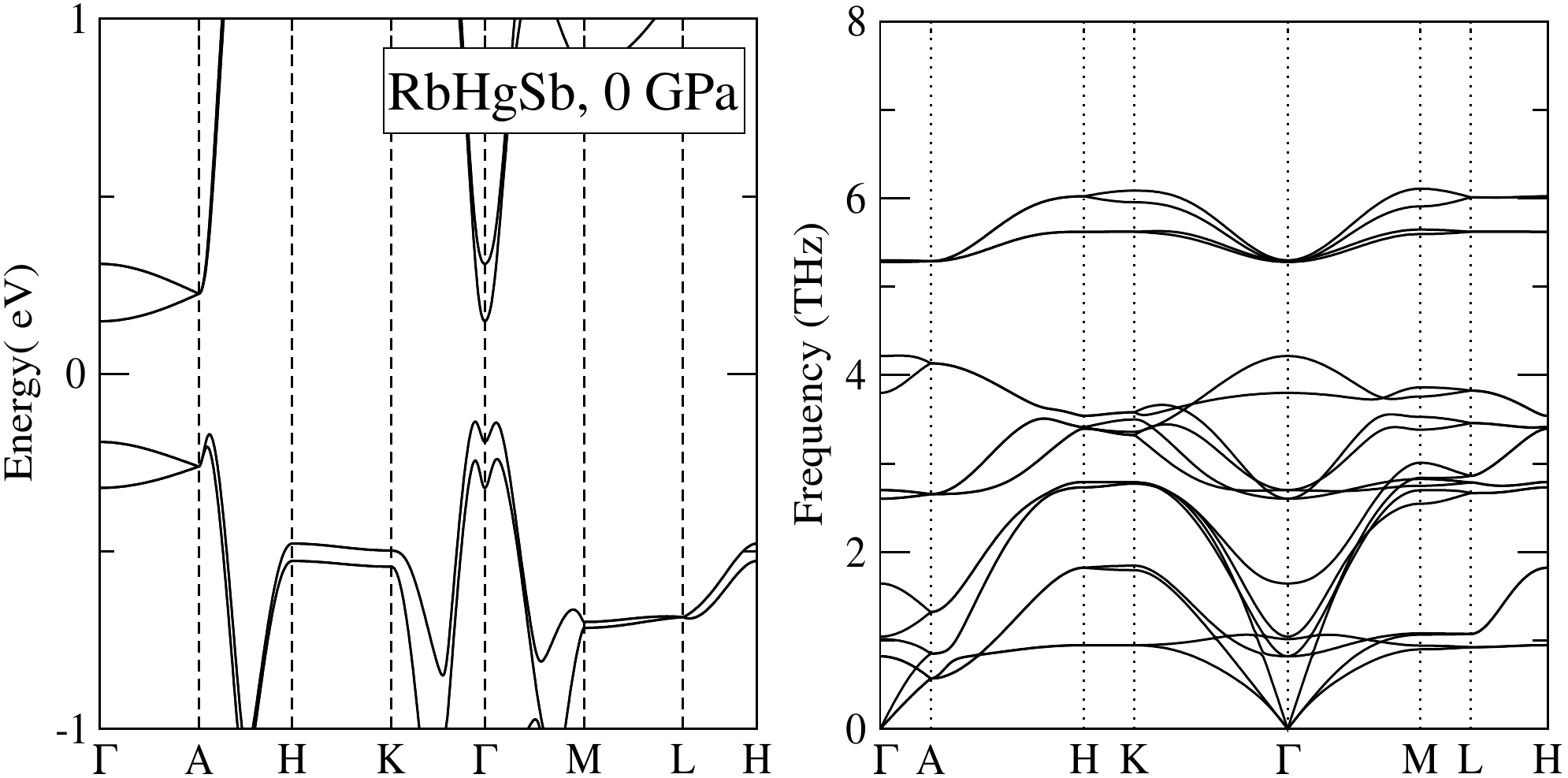}
\includegraphics[width=0.44\textwidth]{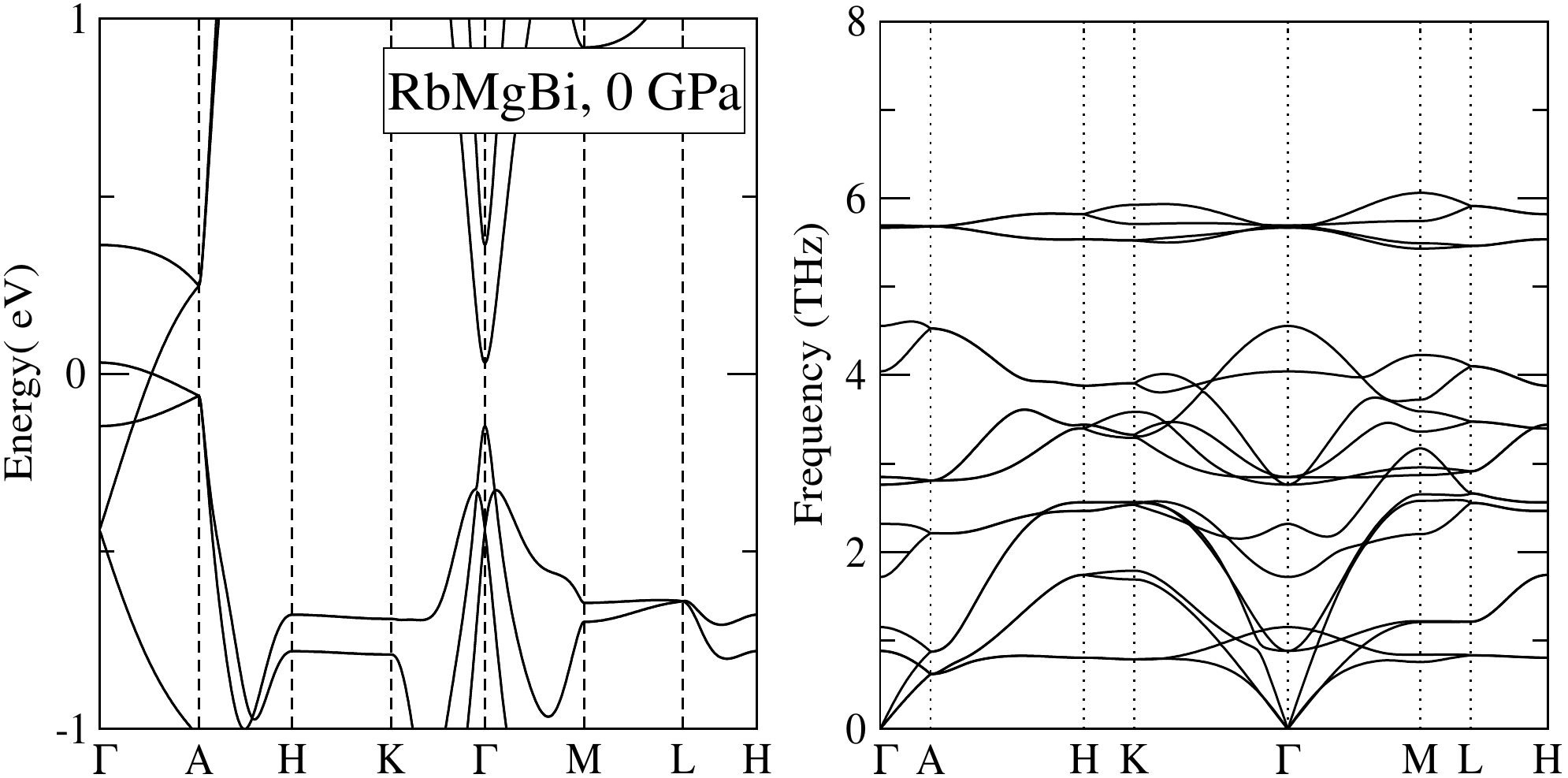}

\label{fig:xyz}
\end{center}
\end{figure*}
\begin{figure*}[!htbp]
\begin{center}
\includegraphics[width=0.44\textwidth]{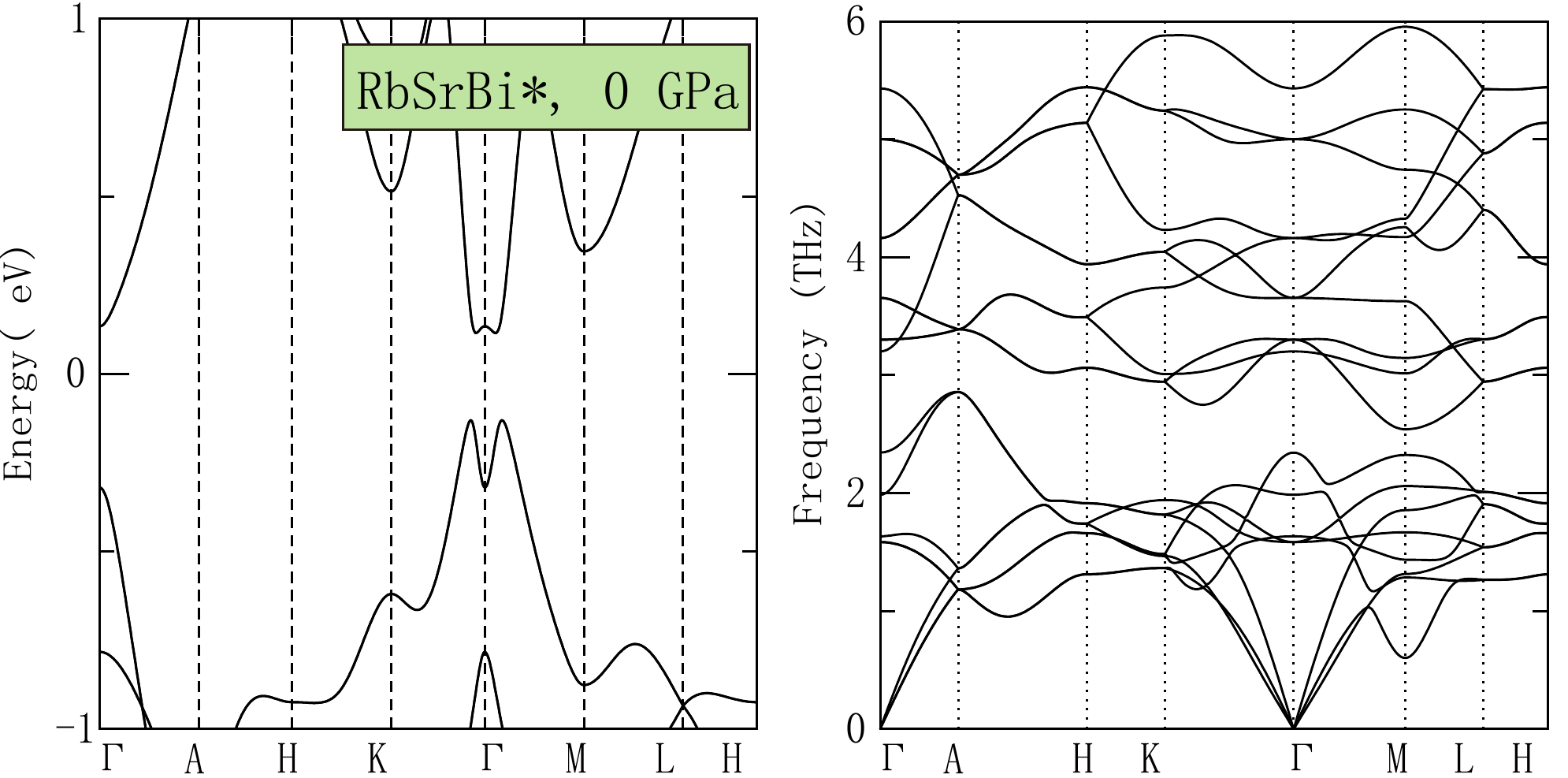}
\includegraphics[width=0.44\textwidth]{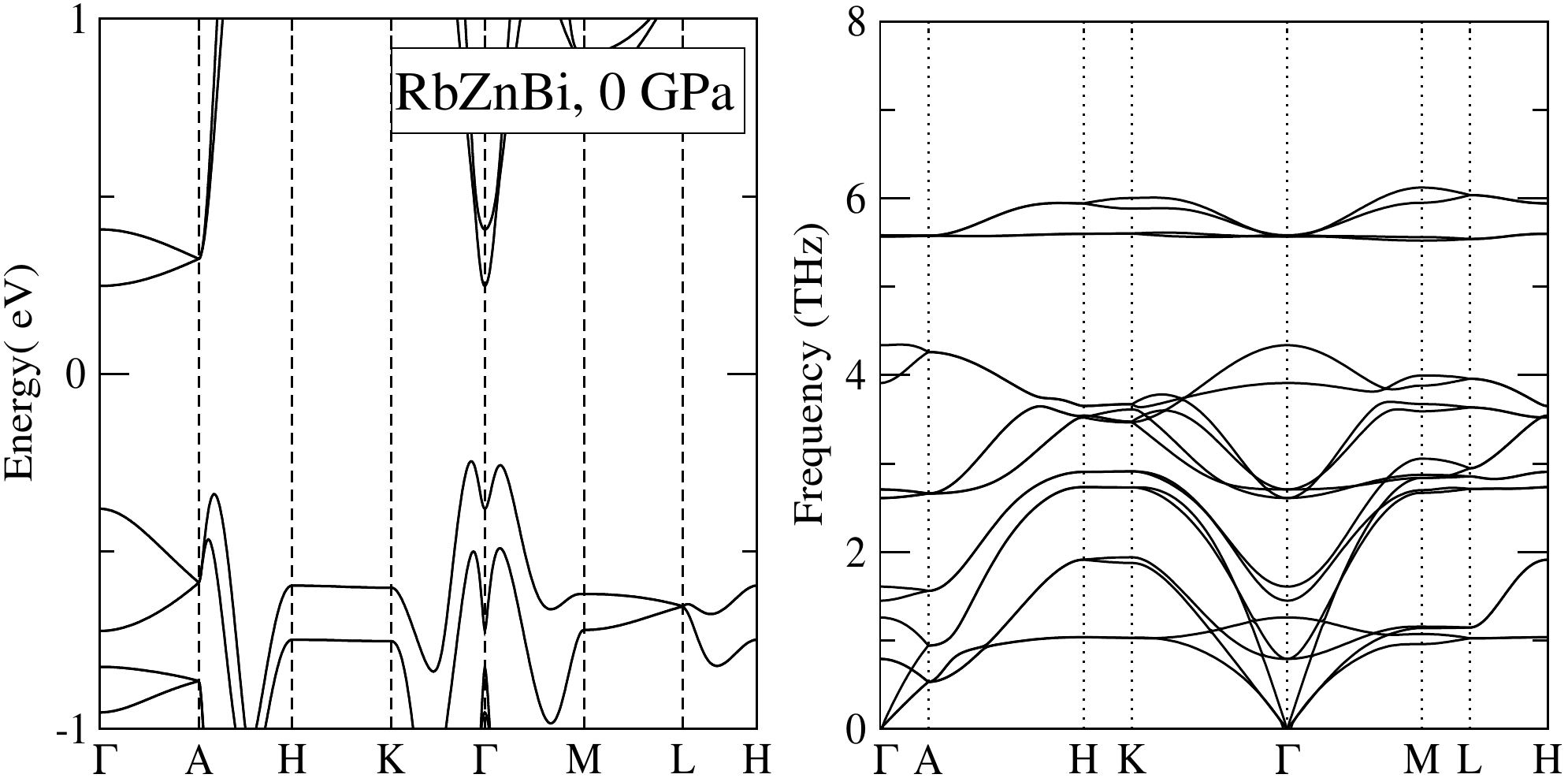}
\includegraphics[width=0.44\textwidth]{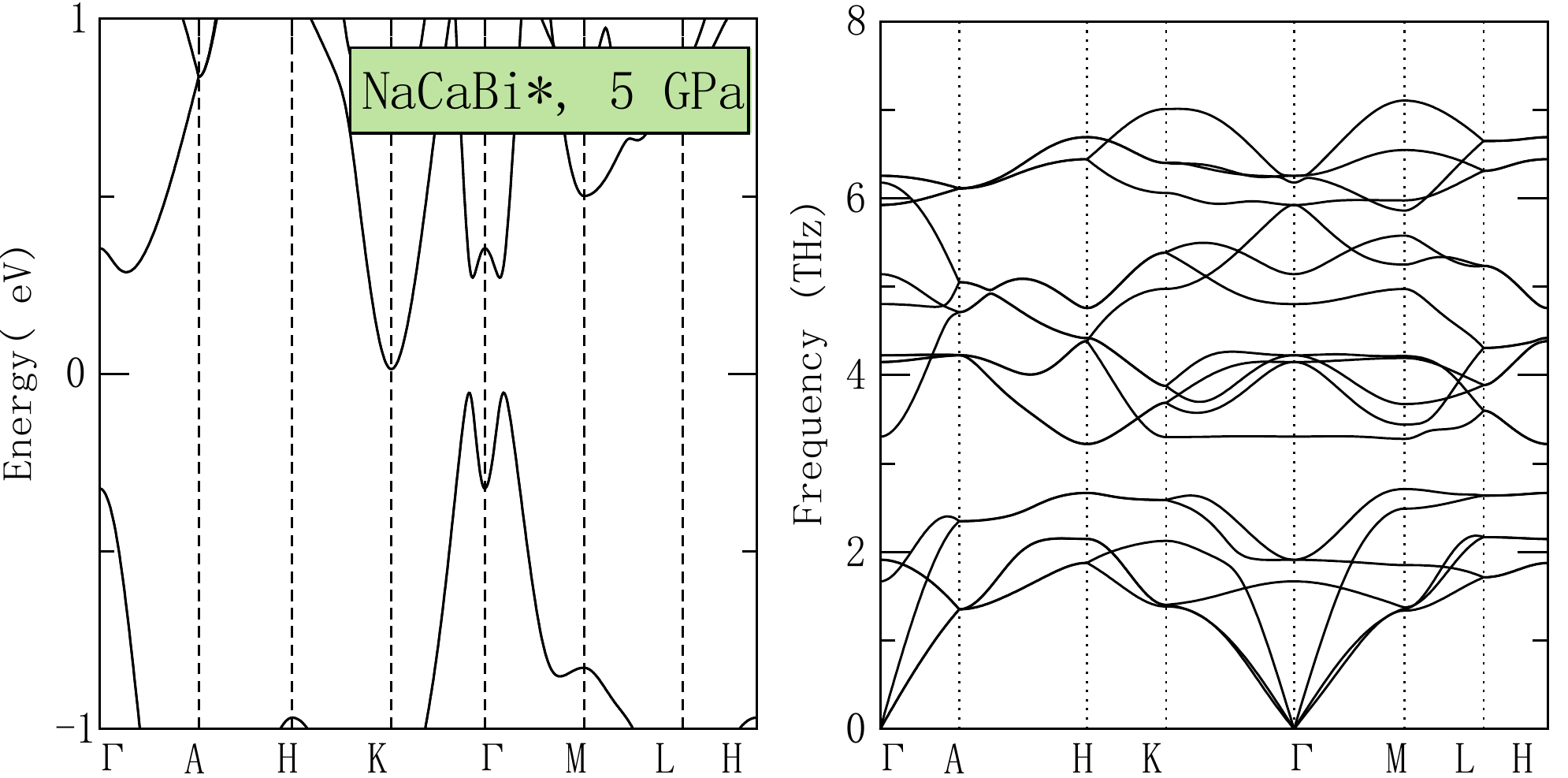}
\includegraphics[width=0.44\textwidth]{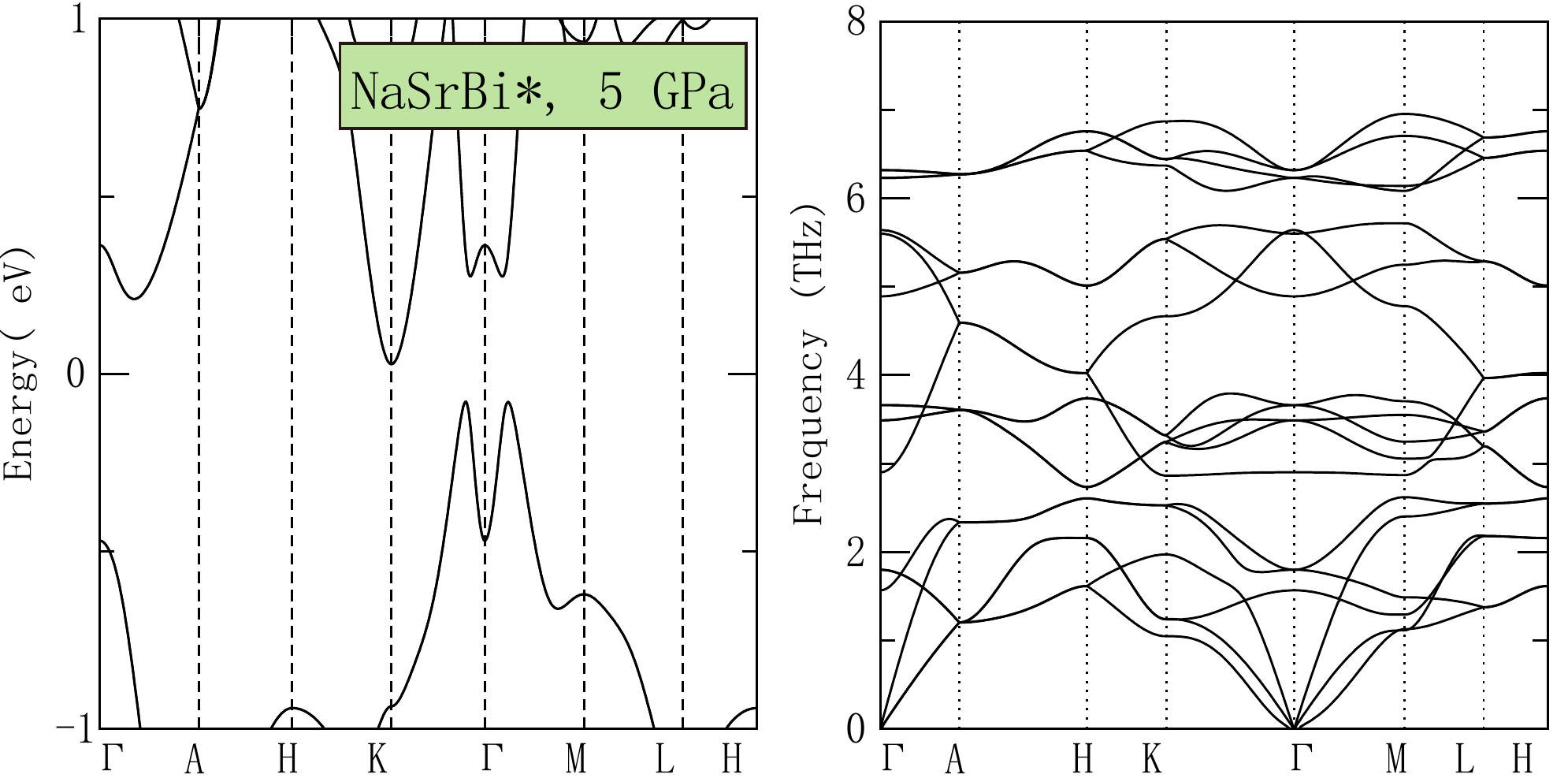}
\includegraphics[width=0.44\textwidth]{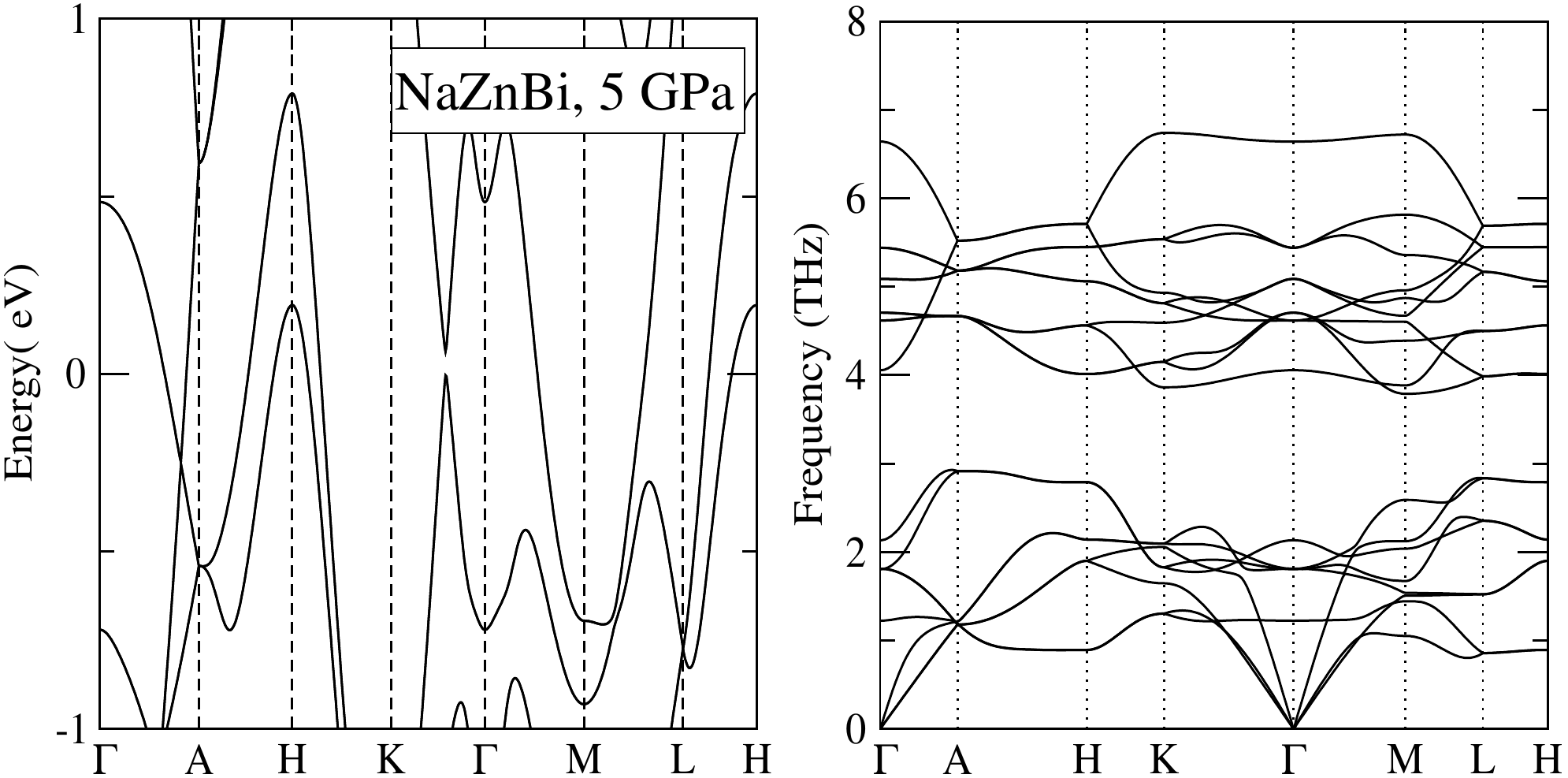}
\includegraphics[width=0.44\textwidth]{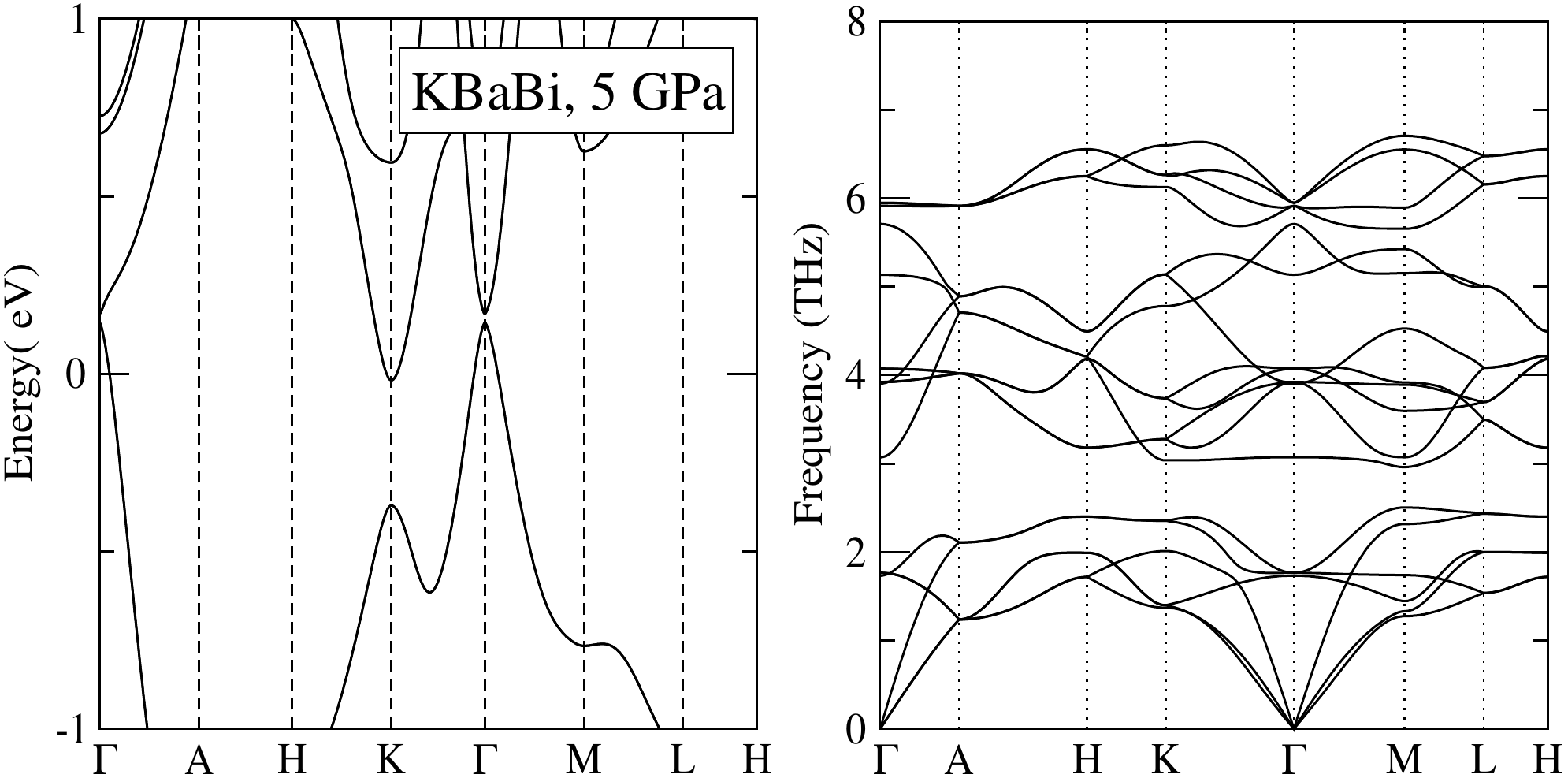}
\includegraphics[width=0.44\textwidth]{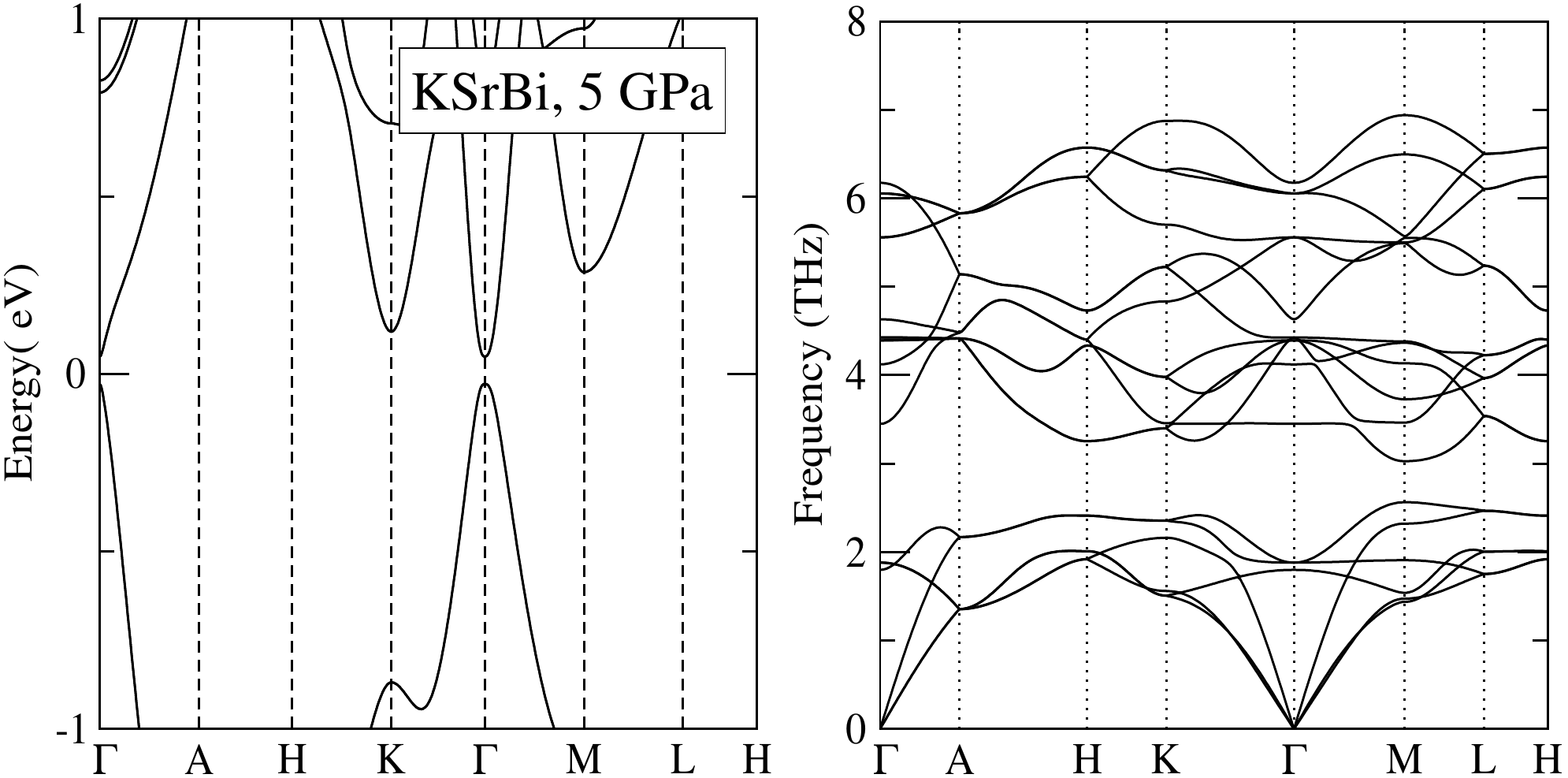}
\includegraphics[width=0.44\textwidth]{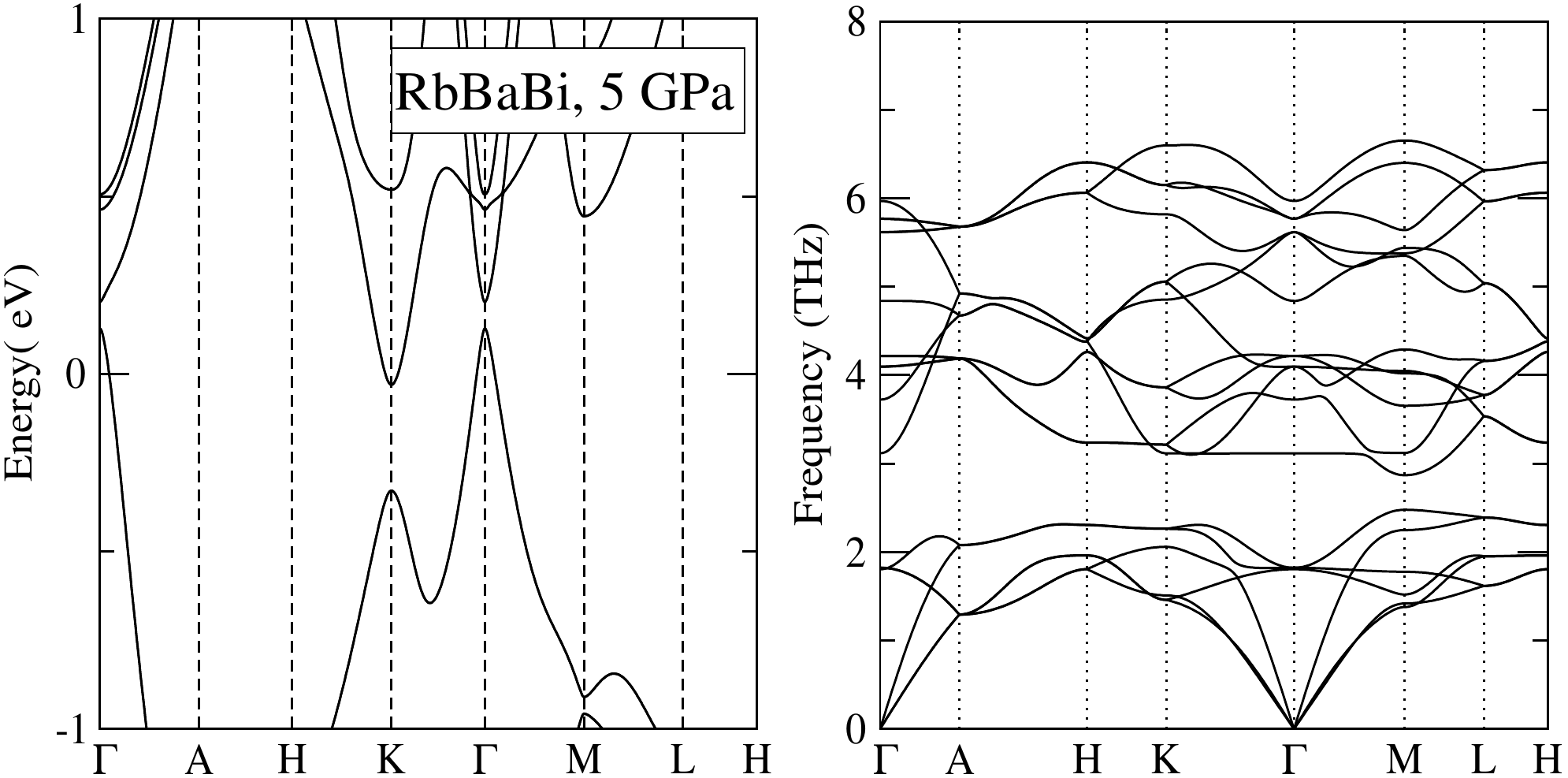}
\includegraphics[width=0.44\textwidth]{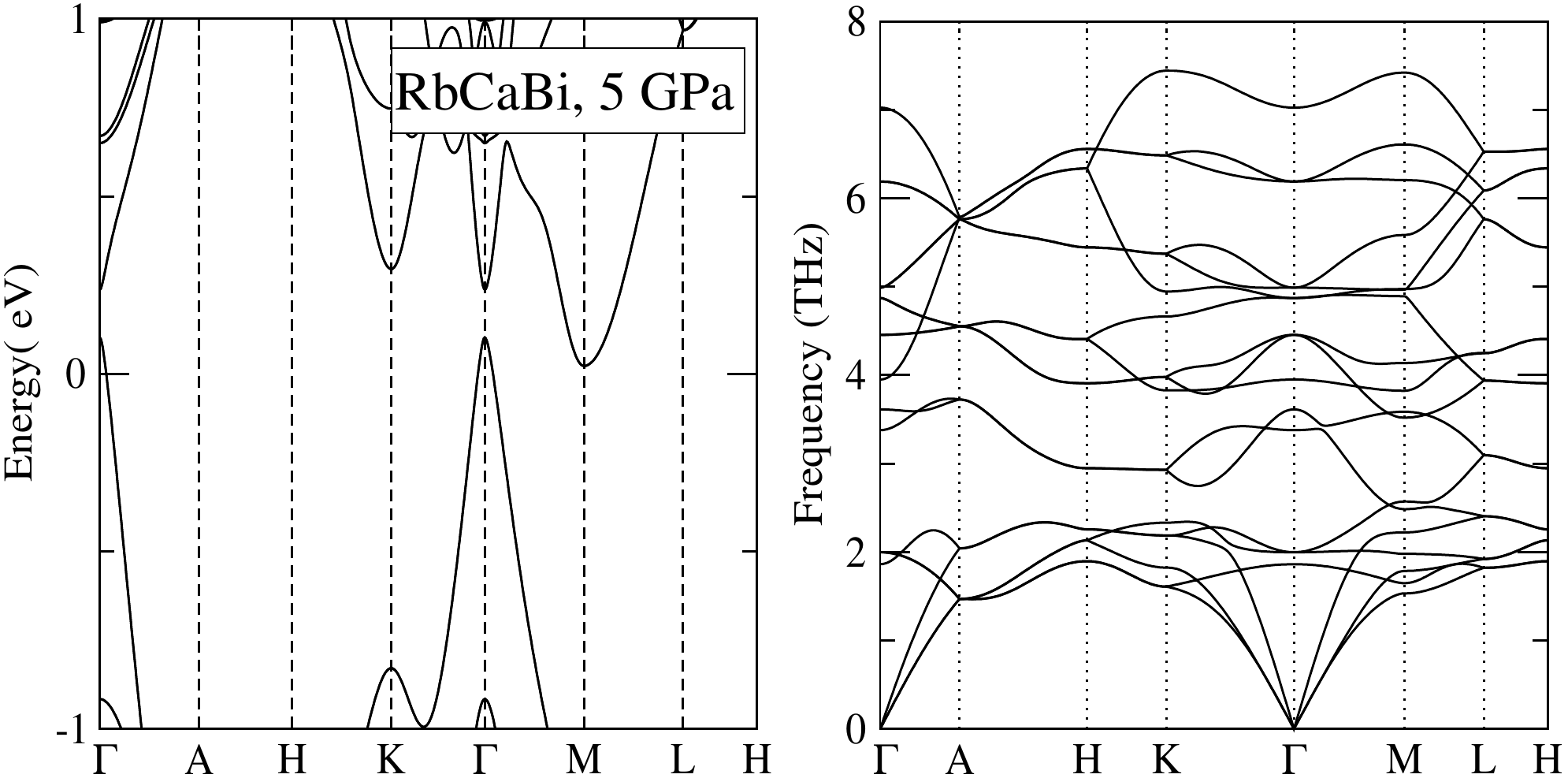}
\includegraphics[width=0.44\textwidth]{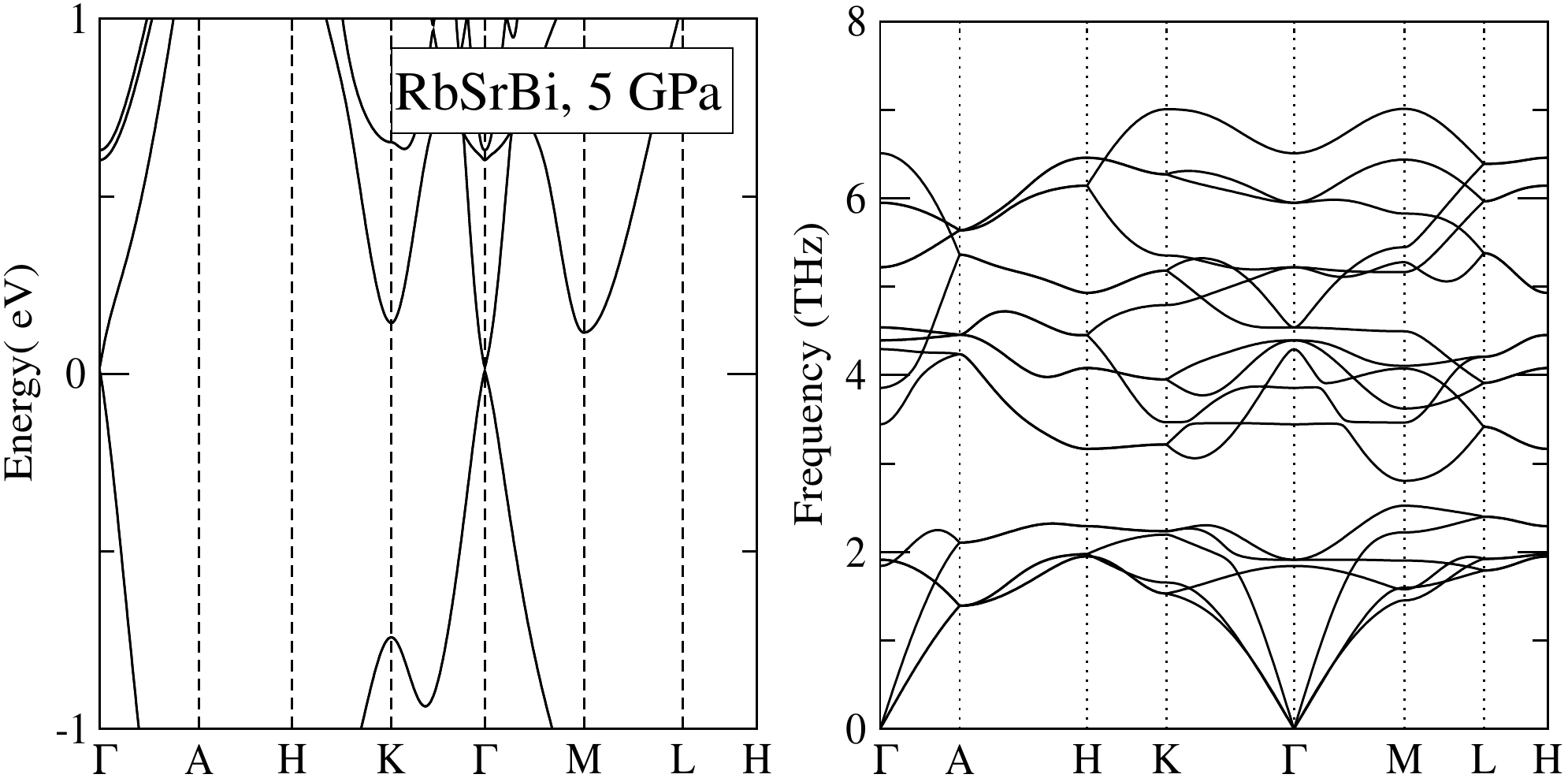}
\includegraphics[width=0.44\textwidth]{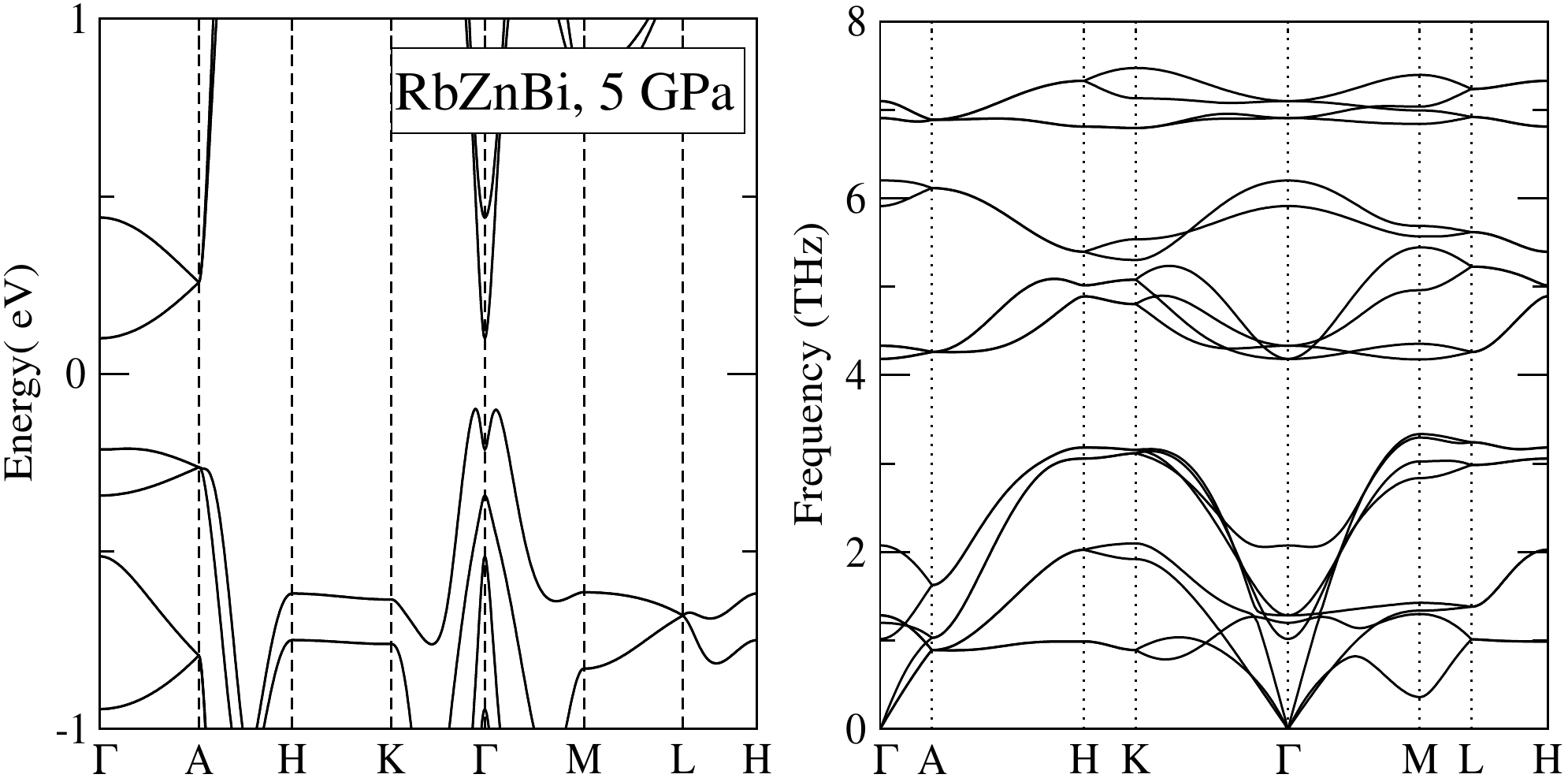}
\includegraphics[width=0.44\textwidth]{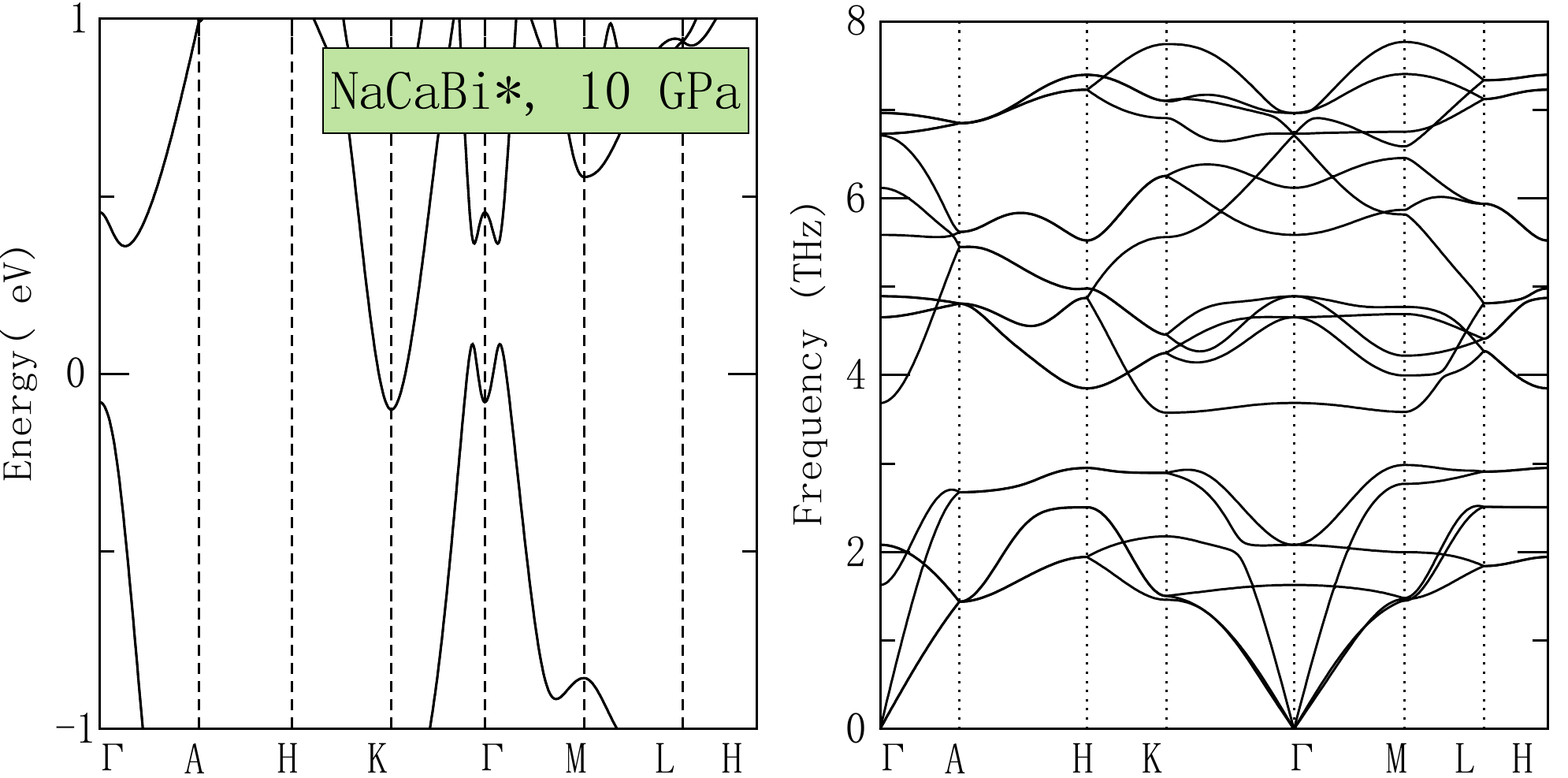}

\label{fig:xyz}
\end{center}
\end{figure*}
\begin{figure*}[!htbp]
\begin{center}
\includegraphics[width=0.44\textwidth]{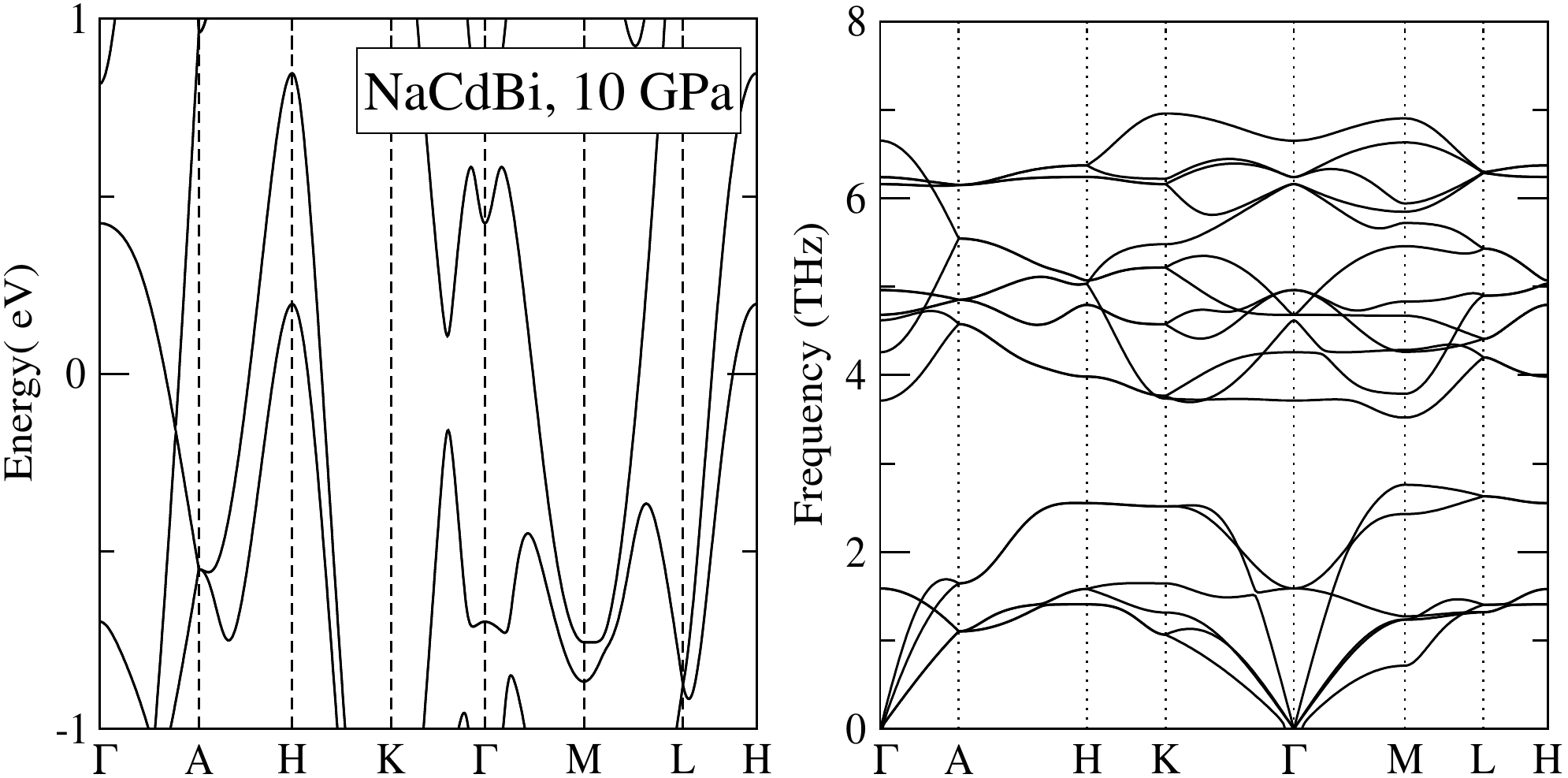}
\includegraphics[width=0.44\textwidth]{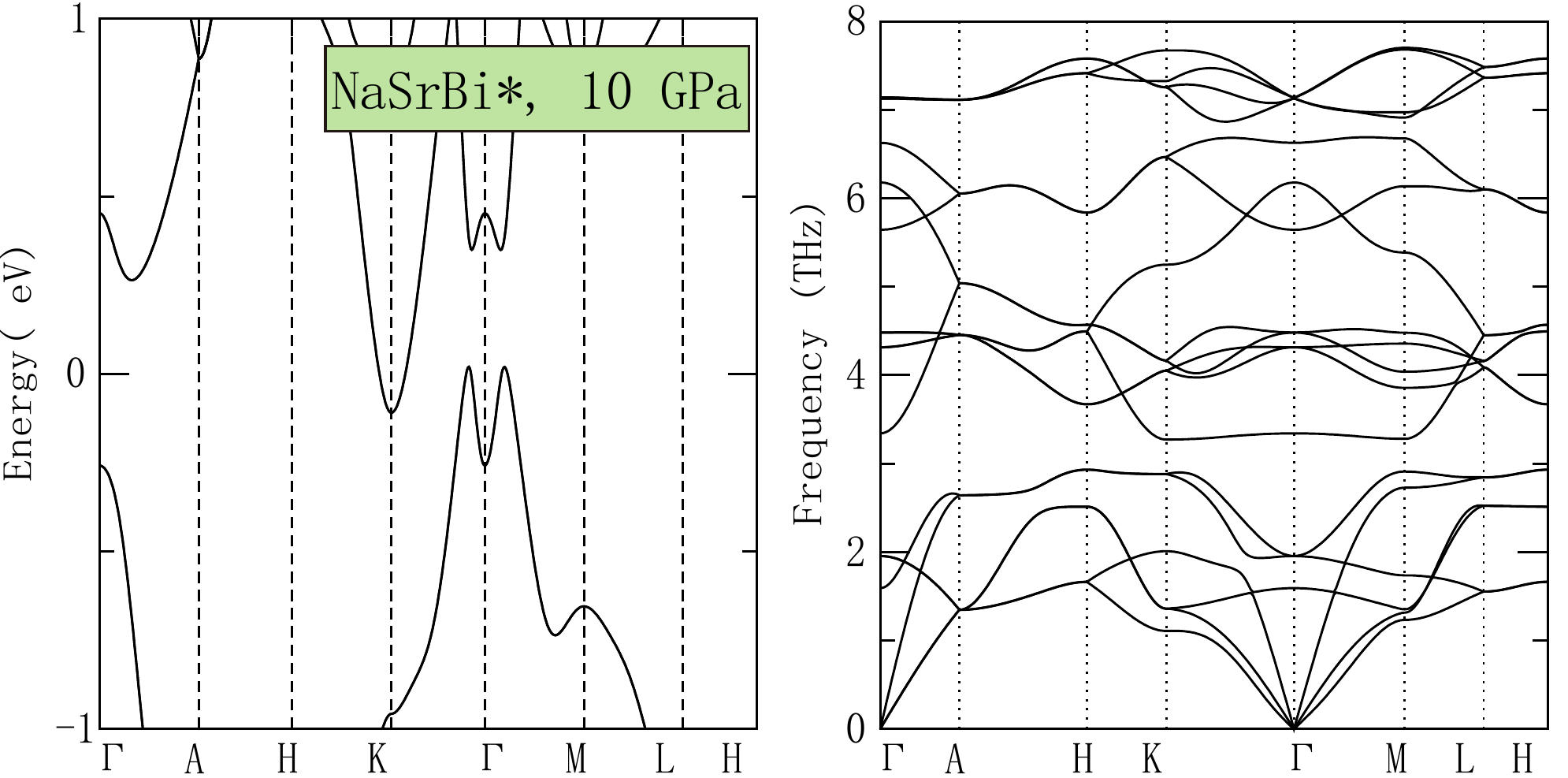}
\includegraphics[width=0.44\textwidth]{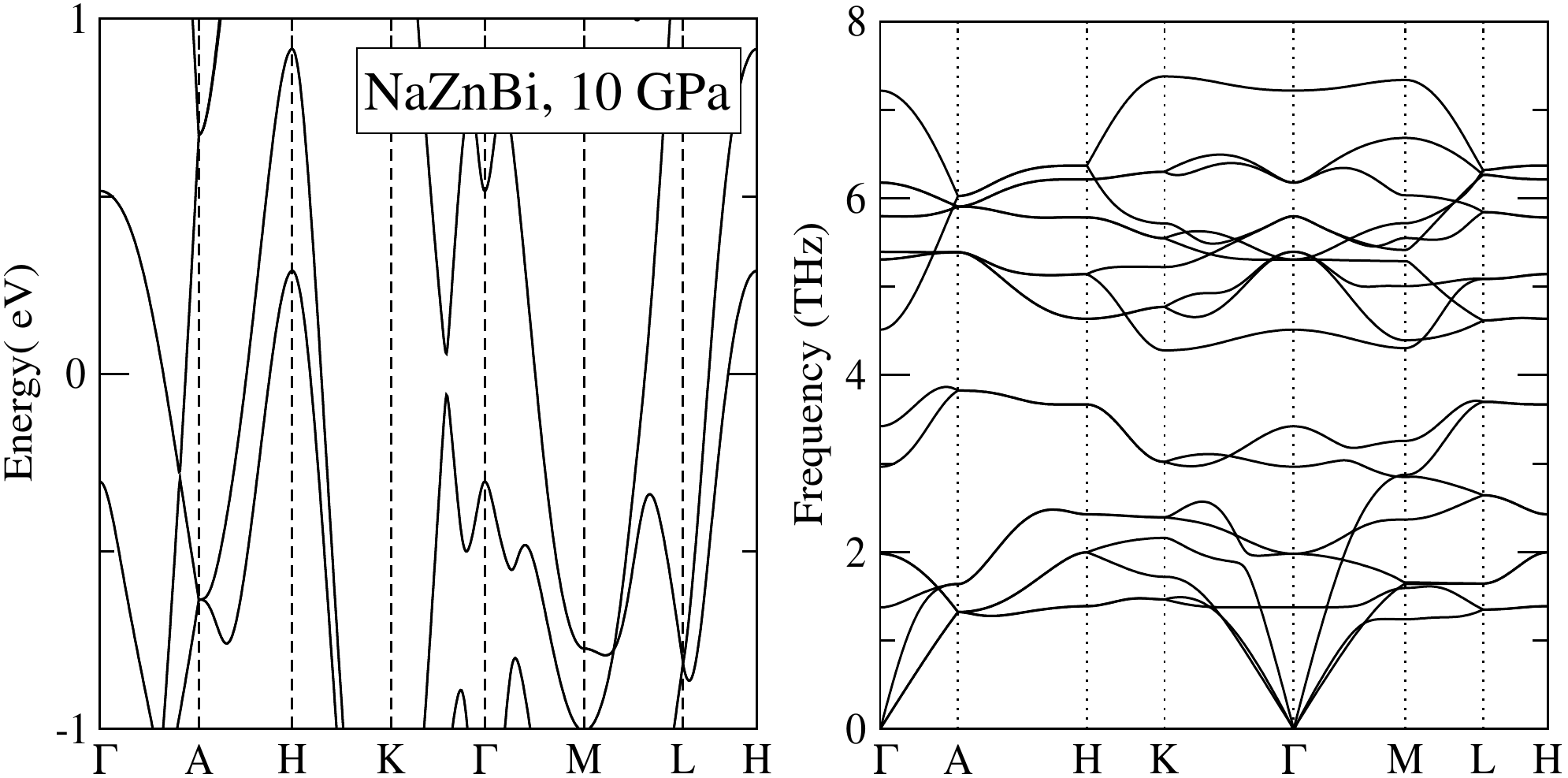}
\includegraphics[width=0.44\textwidth]{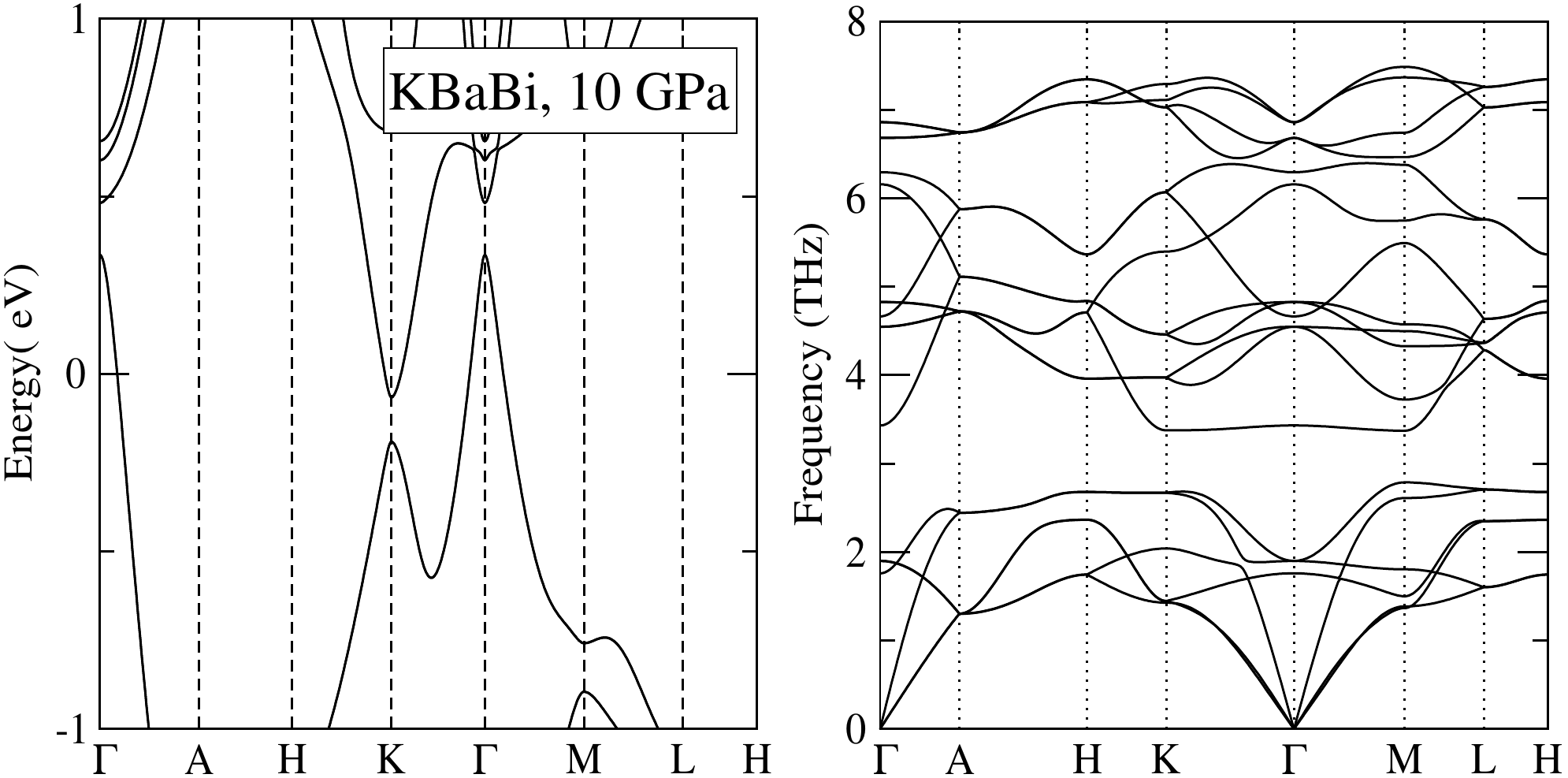}
\includegraphics[width=0.44\textwidth]{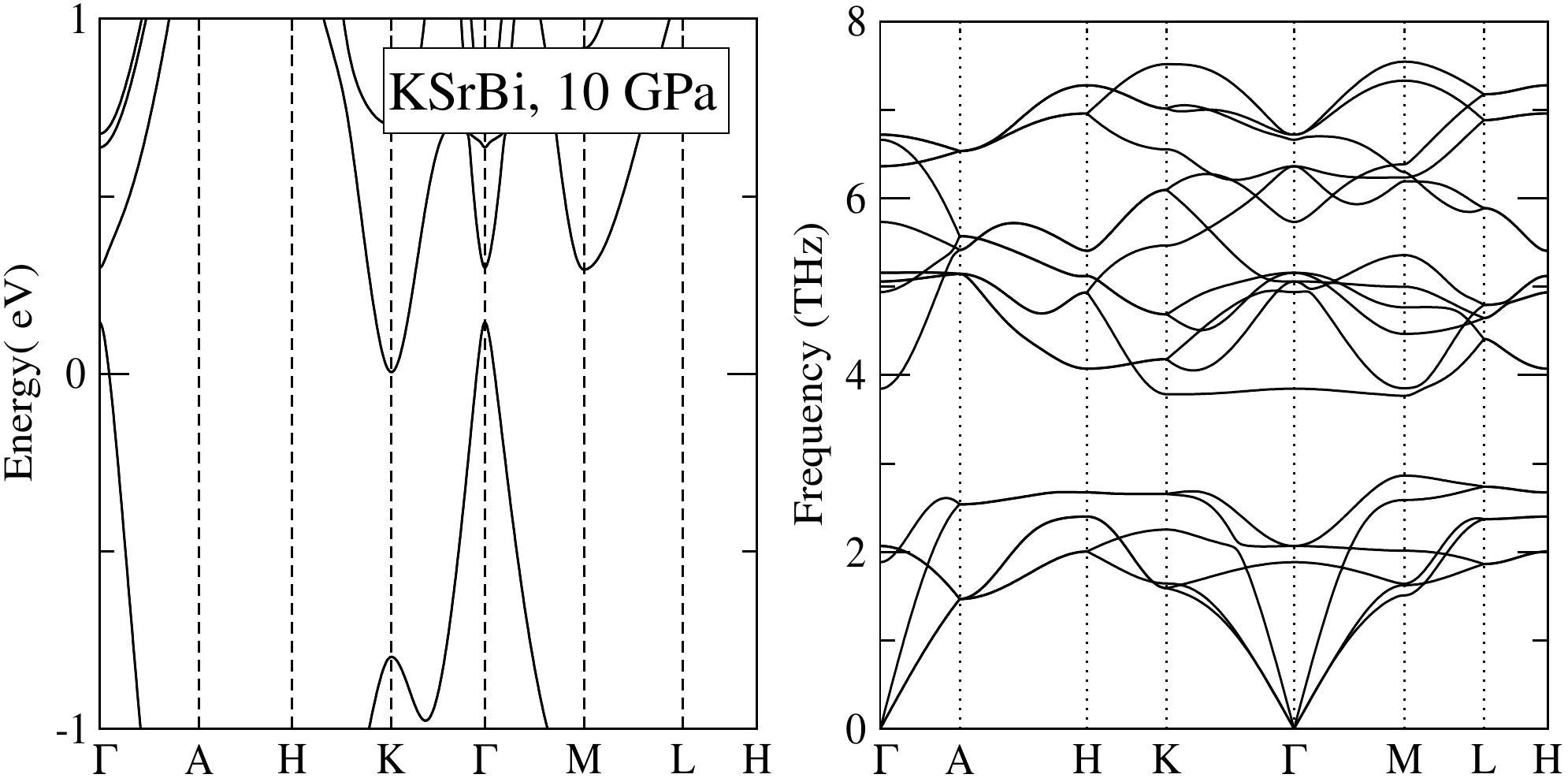}
\includegraphics[width=0.44\textwidth]{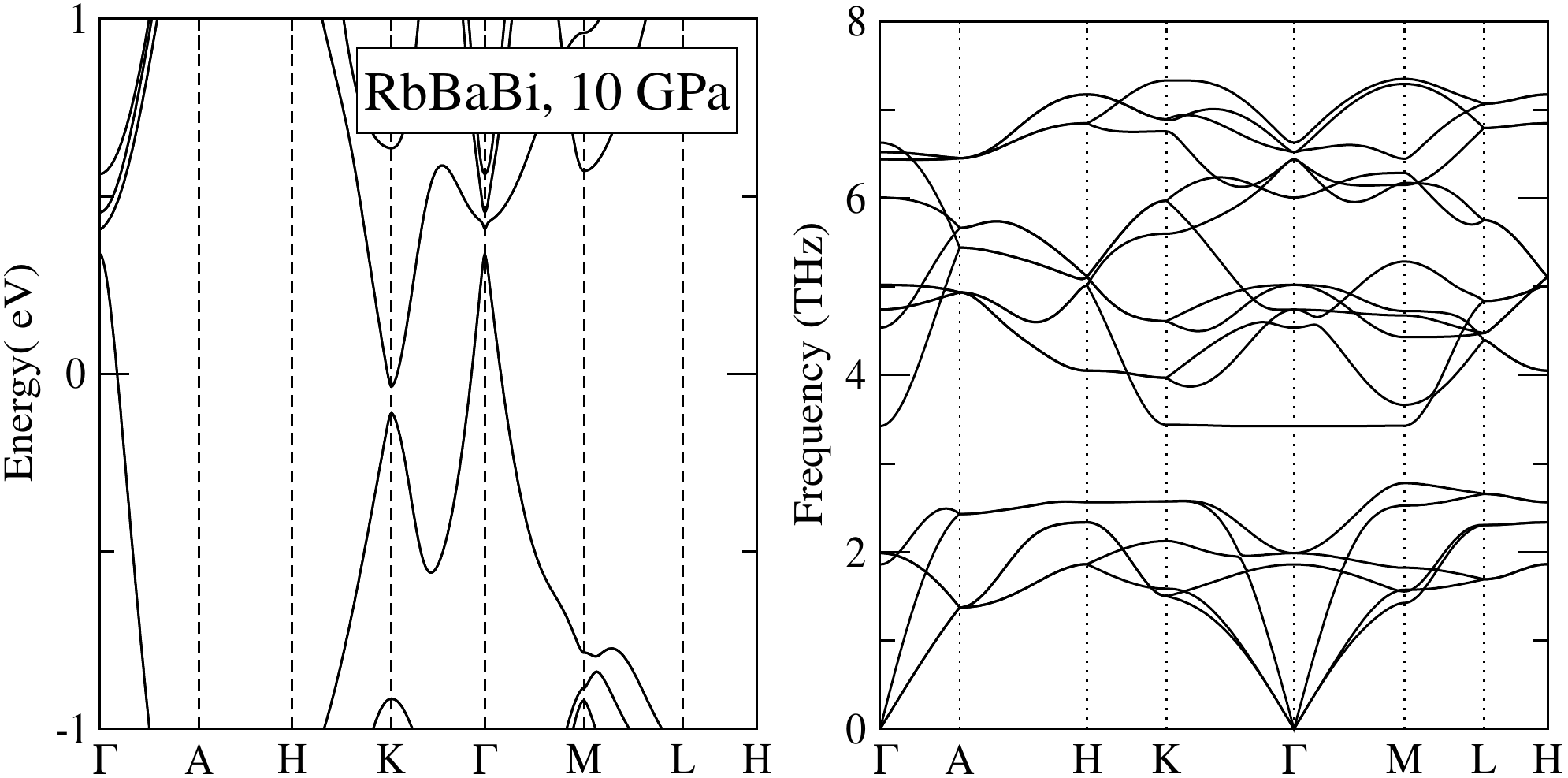}
\includegraphics[width=0.44\textwidth]{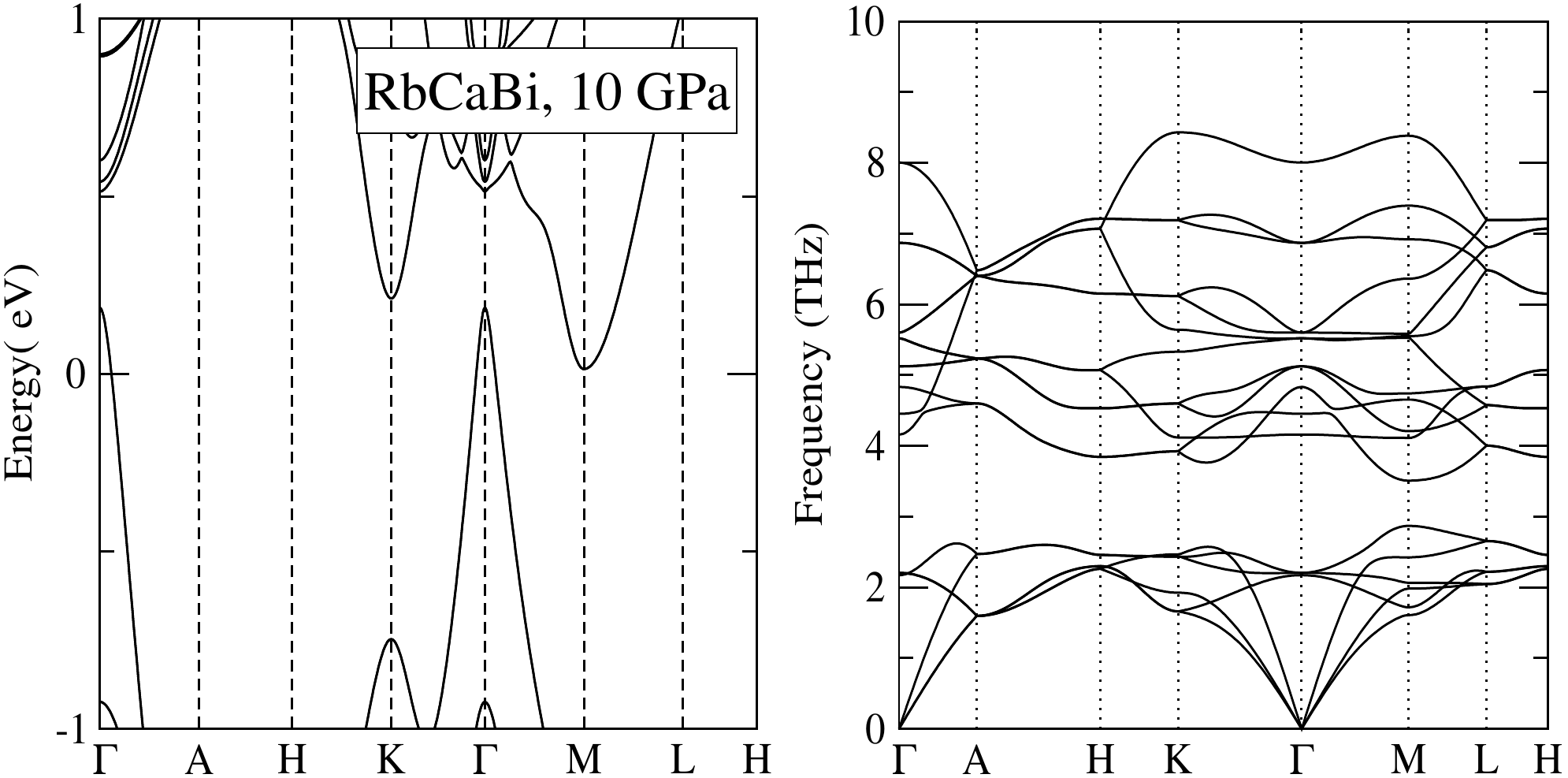}
\includegraphics[width=0.44\textwidth]{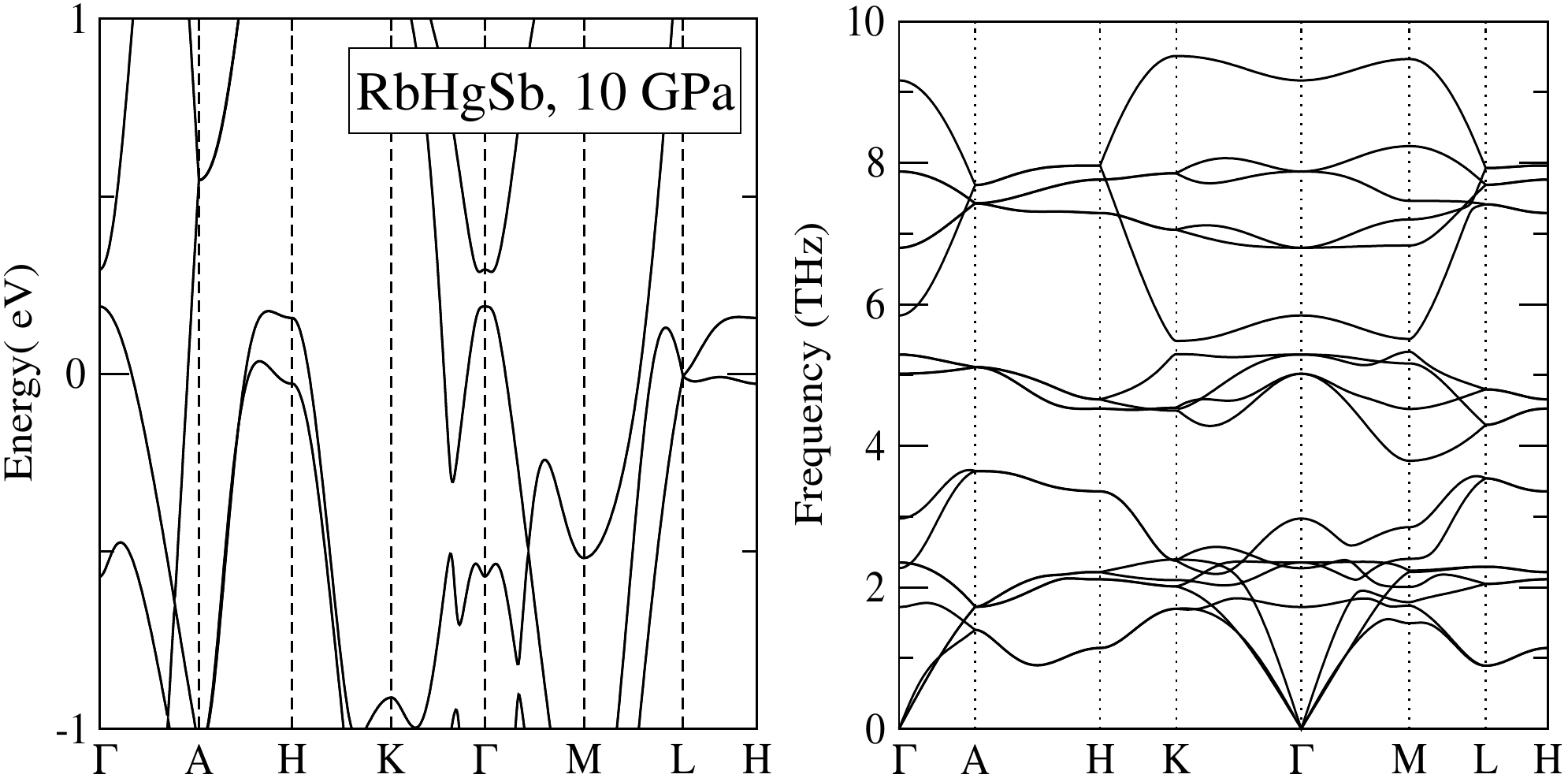}
\includegraphics[width=0.44\textwidth]{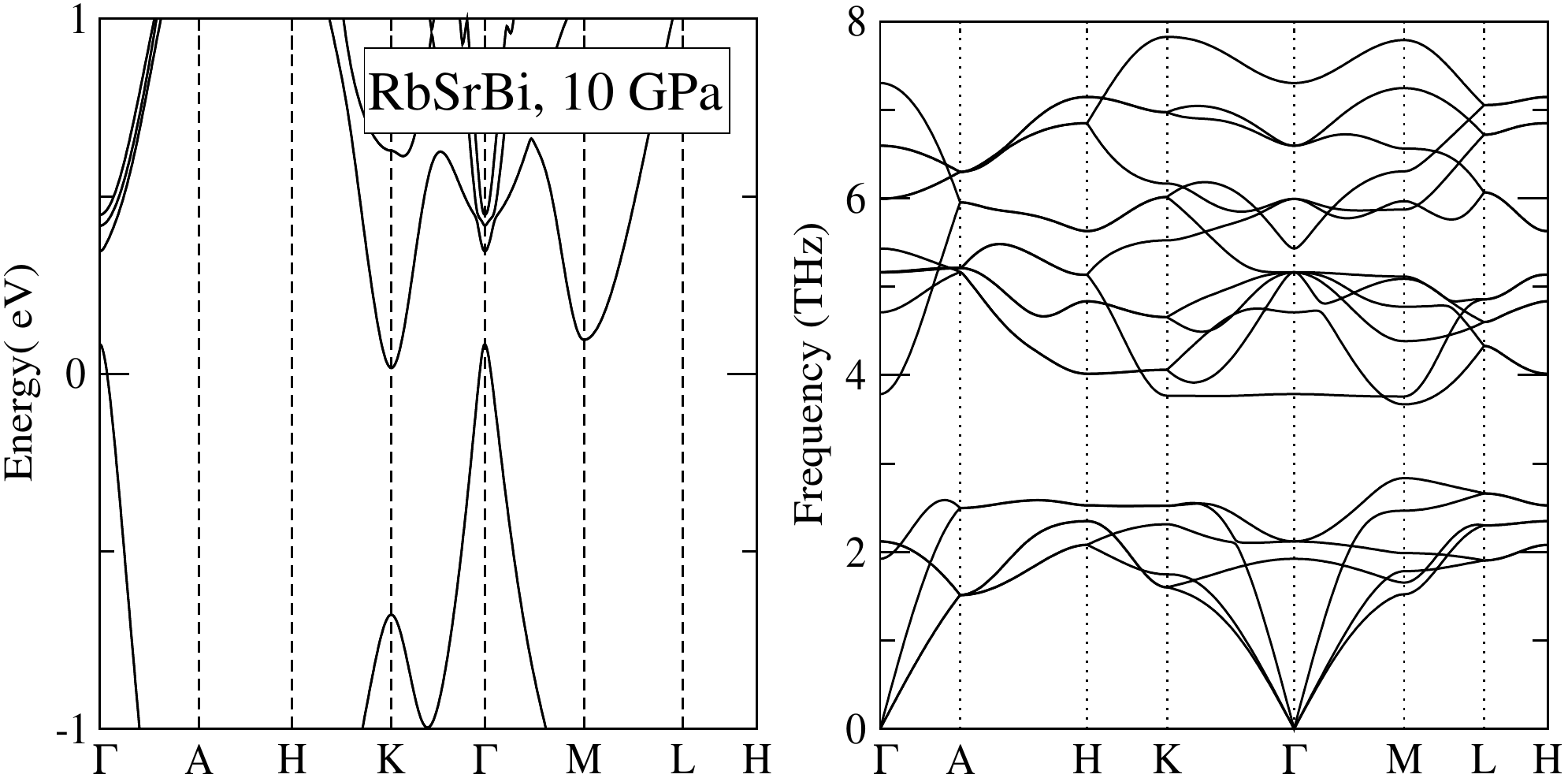}
\caption{%
Band structures and phonon spectra of $P6_{3}/mmc$ RbHgSb and XYBi when they are dynamically stable under 0 GPa, 5 GPa and 10 GPa, respectively. Our calculations reproduce the Dirac point in the $P6_{3}/mmc$ RbMgBi system~\cite{Le2017}.
}
\label{fig:xyz}
\end{center}
\end{figure*}

\begin{table*}[!htp]
\centering
\caption{Dynamically stable RbHgSb and XYBi systems at 0 GPa, 5 GPa and 10 GPa, respectively.}
   \begin{tabular}{l r}%
       \hline\hline
          \textbf{Pressure(GPa)}       & \textbf{RbHgSb and XYBi systems}                 \\
            \hline
            0 GPa    &  NaCaBi*,KBaBi*,KSrBi*,KZnBi,RbBaBi*,RbCdBi,RbHgSb,RbMgBi,RbSrBi*,RbZnBi  \\
            5 GPa    &  NaCaBi*,NaSrBi*,NaZnBi,KBaBi,KSrBi,RbBaBi,RbCaBi,RbSrBi,RbZnBi   \\
            10 GPa   &  NaCaBi*,NaCdBi,NaSrBi*,NaZnBi,KBaBi,KSrBi,RbBaBi,RbCaBi,RbHgSb,RbSrBi             \\
       \hline\hline
  \end{tabular}
    \begin{tablenotes}
        \footnotesize
        \item[1] Structures with ``*'' behind denote those possessing similar band structures with $P6_{3}/mmc$ NaCaBi.
      \end{tablenotes}
\label{table:xyz}
\end{table*}

According to the online Materials Project~\cite{MaterProject} database, the Inorganic Crystal Structure Database~\cite{ICSD} and the earlier work~\cite{Zhang2012}, RbHgSb and several XYBi systems with X = \{Na, K,Rb\} and Y = \{Mg, Ca, Sr, Ba, Zn, Cd, Hg\} can crystallise into the ZrBeSi-type structure of $P6_{3}/mmc$. We performed calculations in terms of the lattice dynamical stability for them at 0 GPa, 5 GPa and 10 GPa, respectively. Among, only the dynamically stable ones are listed in Table~\ref{table:xyz}. Their phonon spectra and band structures are giving in Fig~\ref{fig:xyz}.
From Fig~\ref{fig:xyz}, we can find they are mainly classified into three classes: the KHgSb-type TCI phase~\cite{Wang2016Hourglass} (\ie KZnBi, RbCdBi, RbHgSb, RbZnBi), the Na$_3$Bi-type Dirac phase~\cite{Wang2012} (\ie NaZnBi, NaCdBi and RbMgBi) and the STI phase. Since we are focused on the STI phase, several of these dynamically stable compounds of the STI phase, having the similar band structures with $P6_{3}/mmc$ NaCaBi in the main text, are denoted by ``*" symbols.
\subsection*{5. HSE calculations}
\begin{figure}[!hbtp]
\begin{center}
\includegraphics[width=0.5\textwidth]{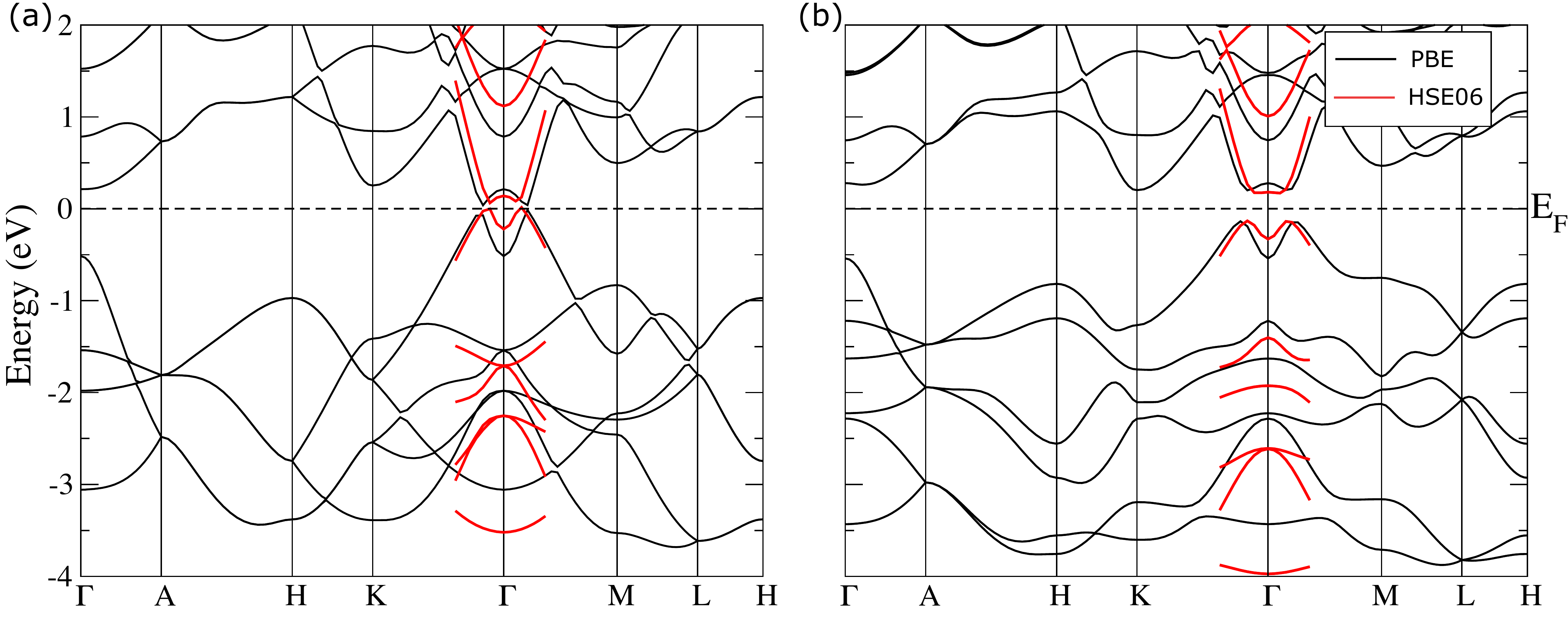}
\caption{%
HSE band structures in comparison with DFT band structures.
}
\label{fig:hse}
\end{center}
\end{figure}

It is known that the PBE functional tends to underestimate the band gap, we recalculated the band structures of $P6_{3}/mmc$ NaCaBi system with the HSE06 functional~\cite{HSE2006}, as shown in Fig.~\ref{fig:hse}. We find the energy dispersions for low energy bands using the HSE06 functional is similar to that using PBE functional, which indicates that the topological nature is solid.

\label{sup:hse}


\subsection*{6. Wannier-based tight-binging model and surface states}
\label{sup:wann}
\begin{figure}[!htbp]
\begin{center}
\includegraphics[width=0.4\textwidth]{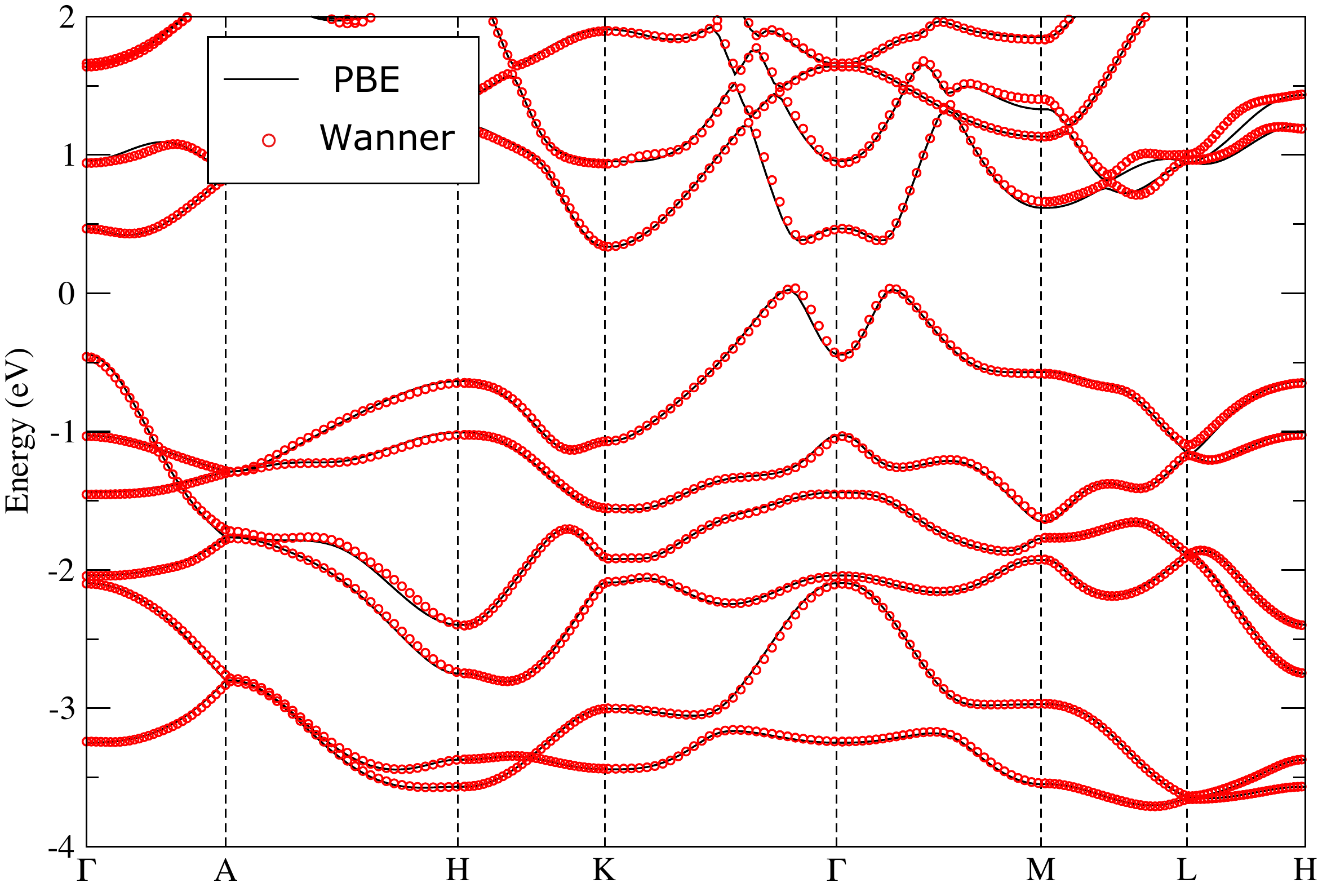}
\caption{%
DFT vs Wannier bands.}
\label{fig:wanfit}
\end{center}
\end{figure}

Surface state plays an important role on justifying/characterizing TIs from the bulk-boundary correspondence. Here we choose Ca-$s$, Ca-$d$ and Bi-$p$ as the projected orbits to build the Wannier-based tight-binging model, it reproduces the DFT band structures well, as shown in Fig.~\ref{fig:wanfit}. Then, we calculated the surface states based on this Wannier-based tight-binging model of NaCaBi on the (010) and (001) surfaces, as presented in Fig.~\ref{fig:surface} in the main text.

\subsection*{7. Matrix presentations of symmetry operators}
\label{sup:kp}
Under the basis of $\Gamma_{7}^{+}$ and $\Gamma_{7}^{-}$, the symmetry operators (\eg $\cal T$, $\cal I$, $C_{6z}$, $C_{2x}$, $C_{2y}$) are represented below:
\beq
\begin{split}
& {\cal T}=\tau_0\otimes (-i)\sigma_y {\cal K};~
  {\cal I}=\tau_z\otimes  \sigma_0;~
    C_{6z}=\tau_0\otimes  e^{-i\frac{\pi}{3}\sigma_z};\\
&C_{2x}=\tau_0\otimes (-i)\sigma_x;~
C_{2y}=\tau_0\otimes (-i)\sigma_y,
\end{split}
\eneq
where ${\cal K}$ denotes the complex conjugation, $\tau_{x,y,z}~(\sigma_{x,y,z})$ are Pauli matrices in the orbital (spin) space. Their (${\cal O}=P, Q, R, \cal T$) representations in momentum space are given by ${\cal O}[k_x,k_y,k_z]^T =g_{\cal O}[k_x,k_y,k_z]^T$
\beq
\begin{split}
&g_{\cal T}=g_{\cal I}=
\left(
\begin{array}{ccc}
-1 & 0 & 0 \\
 0 &-1 & 0 \\
 0 & 0 & 1 \\
\end{array}
\right),~
g_{C6z}=
\left(
\begin{array}{ccc}
 \frac{1}{2} & -\frac{\sqrt{3}}{2} & 0 \\
 \frac{\sqrt{3}}{2} & \frac{1}{2} & 0 \\
 0 & 0 & 1 \\
\end{array}
\right),\\
& g_{C2x}=
\left(
\begin{array}{ccc}
 1 & 0 & 0 \\
 0 & -1& 0 \\
 0 & 0 & -1\\
\end{array}
\right),~
g_{C2y}=
\left(
\begin{array}{ccc}
-1 & 0 & 0 \\
 0 & 1 & 0 \\
 0 & 0 & -1\\
\end{array}
\right).
\end{split}
\eneq
The low-energy effective $\bk\cdot \bp$ Hamiltonian satisfies the condition placed by the symmetry $\cal O$:
\beq
{\cal O}{\cal H}(\delta \bk) {\cal O}^{-1}= {\cal H}(g_{\cal O} \delta \bk)
\eneq
After considering all symmetry restrictions of $D_{6h}$, the low-energy effective model is derived as given in the main text.

\subsection*{8. Pressure, Strain and substrate candidates}
\label{sup:strain}

\begin{figure}[!htbp]
\begin{center}
\includegraphics[width=0.5\textwidth]{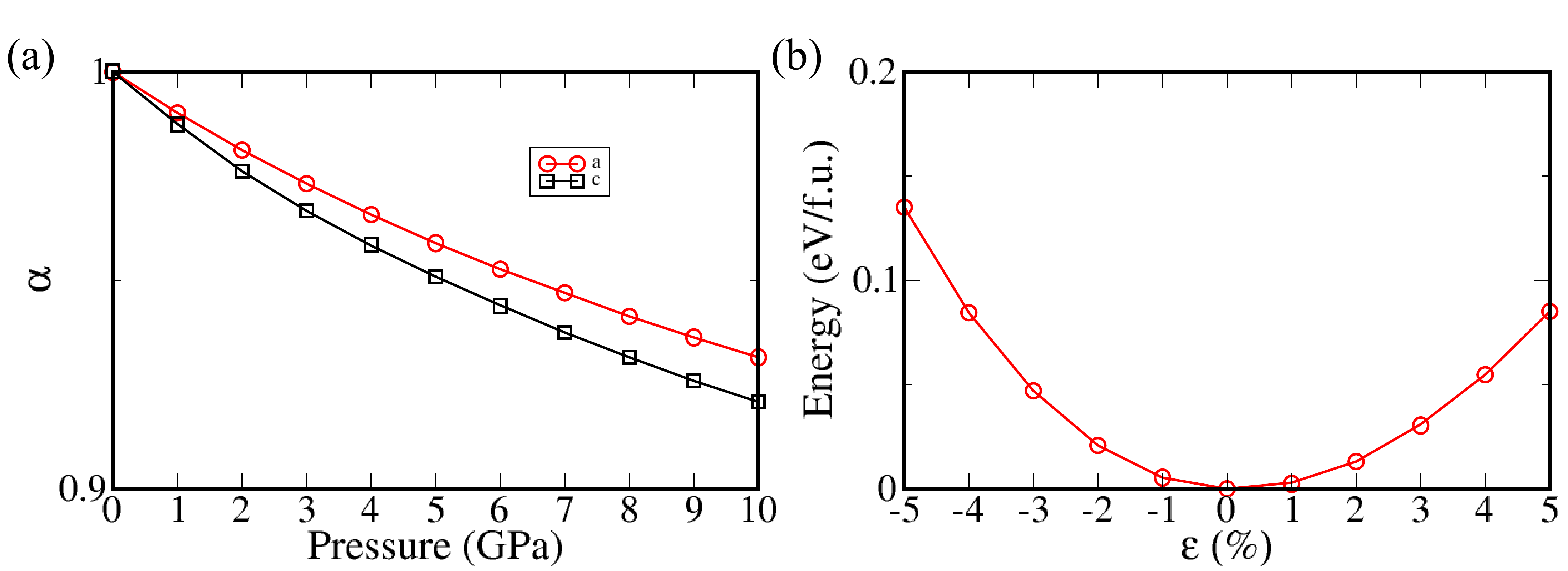}
\caption{%
 (a) Lattice parameters a(= b) and c vs. Pressure. (b) Strain energy vs. the linear biaxial strain along the $x$ and $y$ axes.}
\label{fig:strain}
\end{center}
\end{figure}

From the electronic properties of $P6_{3}/mmc$ NaCaBi at 0 GPa, 5 GPa and 10 GPa shown in Fig.~\ref{fig:xyz}, we find that there is an electron pocket of the conductance band around K dropping to the Fermi level quickly under pressure. And when the applied pressure is as large as 10 GPa, this pocket goes across the Fermi level, which leads $P6_{3}/mmc$
NaCaBi to be a metal. As shown in Fig.~\ref{fig:strain}(a), we can find the lattice parameters a(= b) and c decrease under
pressure. Moreover, the lattice parameters a/c decrease about 1\%/1.3\% at 1 GPa, from which the $P6_{3}/mmc$ phase of
NaCaBi becomes preferred. It should also be noted that though the $P6_{3}/mmc$ NaCaBi changes from an insulator to
a metal from 0 GPa to 10 GPa, the inverted direct gap is always kept. Thus, the STI nature with topological surface
states is kept up to 10 GPa.

Looking along the $c$ axis, the $P6_{3}/mmc$ NaCaBi is a layered hexagonal structure with its layer lying in the xy plane.
Thus, we can introduce substrates perpendicular to the layer plane. As shown in Table~\ref{table:substrates}, we have introduced 31 substrates with the lattice mismatch less than 1\% along the $x$ and $y$ axes to assist growth teams attempting to synthesize the $P6_{3}/mmc$ phase of NaCaBi.
Some commonly used crystals such as KTiO$_3$ and BaNiO$_3$ only have lattice mismatch with NaCaBi of about -0.1\% and 0.3\%, which are very small and can be good substrate candidates. However, it should be noted that, as theoreticians, our proposal is very rough and more complicated experimental
details should be considered before real experiments.

Similar with applying pressure, imposing various strains are also considered as convenient methods to tune the
electronic properties in the condensed systems. Here we focus on applying biaxial linear strain along $a$ and $b$ axes which keeps the symmetries of this system unchanged. From Fig.~\ref{fig:strain}(b), we find the strain energy of this kind of biaxial linear strain increase parabolically with $-5\% \leq \varepsilon \leq 5\%$, which indicates this kind
of strain can be seen as elastic strain in a very large range.
Thus, it offers the experimental team large degree of freedom to choose potential substrates candidates.

\begin{table*}[!htbp]
\centering
\caption{ 31 substrate candidates with the lattice mismatch less than 1\% with the $P6_{3}/mmc$ phase of NaCaBi.}
   \begin{tabular}{l c c c c r}%
       \hline\hline
          \textbf{Formula} & \textbf{No-MP} & \textbf{No-ICSD} & \textbf{Band gap}(eV) & a({\AA}) & \textbf{Mismatch}($\epsilon$)                \\
          \hline
KAl(MoO$_{4}$)$_{2}$ & mp-19352 & 28018 & 3.890 & 5.545 & -0.4\% \\
AlTl(MoO$_{4}$)$_{2}$ & mp-18733 & 250339 & 3.871 & 5.544 & -0.4\% \\
CsAl(MoO$_{4}$)$_{2}$ & mp-505784 & 280947 & 3.879  & 5.551 & -0.3\% \\
RbFe(MoO$_{4}$)$_{2}$ & mp-563010 & 245666 & 2.572 & 5.596 & 0.5\%  \\
HfPd$_{3}$ & mp-11453 & 638769, 104254 & 0 & 5.595 & 0.5\% \\
KTiO$_{3}$ & mp-1180658 & 94775 & 0 & 5.560 & -0.1\% \\
KAg$_{2}$ & mp-12735 & 150142 & 0 & 5.589 & 0.4\% \\
Na$_{2}$LiAu$_{3}$ & mp-12815 & 152090 & 0 & 5.530 & -0.6\% \\
BaNiO$_{3}$ & mp-19138 & 175, 30661 & 1.542 & 5.580 & 0.3\% \\
Ce$_{2}$In & mp-19733 & 621365, 621356, 102186 & 0 & 5.557 & -0.2\% \\
Sm$_{2}$In & mp-19816 & 59527, 59526, 640547, 640544 & 0 & 5.549 & -0.3\% \\
Pr$_{2}$In & mp-19854 & 640268, 108555 & 0 & 5.558 & -0.1\% \\
Nd$_{2}$In & mp-21295 & 59423, 59422, 59421, 640067 & 0 & 5.589 & 0.4\% \\
CaAl$_{12}$O$_{19}$ & mp-28234 & 34394 & 4.320 & 5.564 & $\approx$ 0\%\\
Rb$_{2}$Te & mp-383 & 55157, 55152, 55160, 55154$\cdots$ & 1.372 & 5.574 & 0.1\% \\
In$_{2}$Pt$_{3}$ & mp-510439 & 640295, 59498, 640287 & 0 & 5.575 & 0.2\% \\
SmOs$_{2}$ & mp-570007 & 150521 & 0 & 5.574 & 0.1\% \\
TaPd$_{3}$ & mp-582360 & 648968 & 0 & 5.520 & -0.8\% \\
PF$_{5}$ & mp-8511 & 62554 & 7.109 & 5.563 & $\approx$0\%\\
Ba$_{3}$BPO$_{3}$ & mp-9712 & 402017 & 1.471 & 5.502 & -1.1\% \\
Ba$_{3}$BAsO$_{3}$ & mp-9793 & 402682 & 1.529 & 5.528 & -0.7\% \\
LaAl$_{2}$Ag$_{3}$ & mp-16766 & 57329 & 0 & 5.565 & $\approx$0\% \\
Na$_{3}$OsO$_{5}$ & mp-555476 & 416999 & 0.113 & 5.553 & -0.2\% \\
Li$_{2}$Ta$_{2}$(OF$_{2}$)$_{3}$ &  mp-561011 & 405777 & 3.835 & 5.581 & 0.3\% \\
Sr$_{2}$MgH$_{6}$ & mp-644225 & 88205 & 2.758 & 5.546 & -0.4\% \\
LaBPt$_{2}$ & mp-31052 & 98425 & 0 & 5.522 & -0.8\% \\
LaB$_{2}$Ir$_{3}$ & mp-10112 & 44422 & 0 & 5.543 & -0.4\% \\
BaPd$_{5}$ & mp-2606 & 58672, 616030 & 0 & 5.540 & -0.5\% \\
NaBPt$_{3}$ & mp-28614 & 68092 & 0 & 5.547 & -0.3\% \\
CaB$_{2}$Rh$_{3}$ & mp-28705 & 66767 & 0 & 5.551 & -0.3\% \\
SrZn$_{5}$ & mp-638 & 106114, 418615 & 0 & 5.541 & -0.4\% \\
\hline\hline
  \end{tabular}
    \begin{tablenotes}
        \footnotesize
        \item[1] 1. ¡°No-MP¡± column denotes the materials ID recorded in the Materials Project, while ¡°No-ICSD¡± column denotes the materials ID recorded in the ICSD.
        \item[2] 2. $\epsilon > 0$ denotes the lattice parameter a of the substrate candidate is larger than that of $P6_{3}/mmc$ NaCaBi, while $\epsilon < 0$ denotes the
lattice parameter a of the substrate candidate is smaller than that of $P6_{3}/mmc$ NaCaBi.
      \end{tablenotes}
\label{table:substrates}
\end{table*}
\clearpage
\bibliographystyle{naturemag}

\end{document}